\documentclass [a4paper,12pt]{article}
\usepackage[english]{babel}
\usepackage{graphicx,amsmath,amsfonts,amssymb,slashed,upgreek,color,caption,amscd,cite,cancel}

\newcommand{\be}{\begin{equation}}
\newcommand{\ee}{\end{equation}}
\newcommand{\bea}{\begin{eqnarray}}
\newcommand{\eea}{\end{eqnarray}}

\newcommand{\PP}{\Psi}
\newcommand{\F}{\Phi}
\newcommand{\hg}{\hat g}
\newcommand{\nablav}{{\nabla}}

\usepackage[stable]{footmisc}

\def\MM{M_{*}}

\begin{document}

\begin{center}
\Large{\textbf{The Effective Field Theory of Dark Energy}} \\[0.5cm]
 
\large{Giulia Gubitosi$^{\rm a}$, Federico Piazza$^{\rm b}$ and Filippo Vernizzi$^{\rm a}$}
\\[0.5cm]

\small{
\textit{$^{\rm a}$ CEA, IPhT, 91191 Gif-sur-Yvette c\'edex, France\\ CNRS, URA-2306, 91191 Gif-sur-Yvette c\'edex, France}}

\vspace{.2cm}

\small{
\textit{$^{\rm b}$ Paris Center for Cosmological Physics (PCCP) and Laboratoire APC,\\  Universit\'e Paris 7, 75205 Paris, France
}}

\end{center}

\vspace{2cm}

\begin{abstract}
We propose a universal description of dark energy and modified gravity that includes all single-field  models. 
By extending a formalism previously applied to inflation, 
we consider the metric universally coupled to matter fields and we write  in terms of it the most general unitary gauge action consistent with the residual unbroken symmetries of spatial diffeomorphisms.  
Our action is particularly suited for cosmological perturbation theory:  the background evolution depends on only three operators. All other operators start at least at quadratic order in the perturbations and their effects can be studied independently and systematically. In particular, we focus on the properties of a few operators which  appear in non-minimally coupled scalar-tensor gravity and galileon theories. In this context, we study the mixing between gravity and the scalar degree of freedom. We assess the quantum and classical stability, derive the speed of sound of fluctuations and the renormalization of the Newton constant. The scalar can always be de-mixed from gravity at quadratic order in the perturbations, but not necessarily through a conformal rescaling of the metric. 
We show how to express covariant field-operators in our formalism and give several explicit examples of dark energy and modified gravity models  in our language. Finally, we discuss the relation with the covariant EFT methods recently appeared in the literature.

\end{abstract}

\newpage 
\tableofcontents

\vspace{.5cm}

\section{Introduction}

The discovery of cosmic acceleration has motivated a tremendous amount of observational and theoretical activity. On the observational side, a wealth of precious cosmological data is awaiting us, thanks to the next generation of large scale structure surveys such as EUCLID~\cite{euclid1,euclid2} and BigBOSS~\cite{bigboss}. By accessing a very large number of modes, such experiments will be sensitive to the details of structure formation and therefore offer the unique  possibility of discriminating between competing models of dark energy and modified gravity (hereafter, in short, DE) on the basis of  dynamical and clustering properties~\cite{Ishak:2005zs}. On the theoretical side, the number and variety of proposed models for the cosmic acceleration (see e.g.~\cite{costas,lucashin} for recent reviews) is by no means less impressive. Such an overwhelming production is certainly a resource for cosmology, but also a somewhat embarrassing load of  material difficult to deal with. 
One reason for concern is the unnatural complexity of many DE models---especially when it is a naturalness problem that they are allegedly addressing. This is of course, to some extent, a subjective issue, difficult to quantify and find general agreement on. If we decide to maintain a democratic view on all different proposals and leave the final verdict to observations we face a more pragmatic problem:  efficiently discriminating between different  models. Each one has a certain number (and type) of parameters and has to be fitted against data independently from the others, by first solving the background and then the perturbation equations. As a matter of fact, in most cases, the performances of each DE model are compared ``one against one" with those of the best contender so far, $\Lambda$CDM.

This situation calls for a ``unifying'' and ``effective'' approach to DE modeling.  ``Unifying'' in the sense that it should incorporate as many different models as possible as special cases; ``effective'', here, in the non-technical sense of being readily testable by observations and not committed to specific DE models, nor to their original motivations. A substantial effort has been put in the recent years in this direction; see for instance\cite{post-fried1,post-fried2,Creminelli:2008wc,PZW,Baker:2011jy,BF,Jimenez:2012jg,Battye:2012eu,Baker:2012zs,Sawicki:2012re}.

Most models of dark energy and modified gravity can be described---in their relevant re\-gimes---with the only addition of a scalar degree of freedom to the Einstein-Hilbert action. This is reminiscent of inflation, where a scalar field is needed to break de Sitter invariance.
In the context of inflation, the goals mentioned above were efficiently achieved  by the `effective field theory (EFT) of inflation', initiated by Creminelli {\it et al.}~in Ref.~\cite{EFT1} and then more systematically developed by  Cheung {\it et al.}~in Ref.~\cite{EFT2}\footnote{This formalism made its first appearance in Ref.~\cite{ghost}, where it was used to study the coupling of the ghost condensate with gravity. See also~\cite{multifield,dissipative} for later generalizations to multi-field inflation and dissipative effects in inflation.}. The idea is to apply EFT directly to cosmological {\it perturbations}, by treating them as the Goldstone boson of spontaneously broken time-translations. The often invoked analogy is  with the spontaneous breaking of the $SU(2) \times U(1)$ gauge symmetry in the Standard Model. In unitary gauge the would-be Goldstone bosons are ``eaten'' by the longitudinal degrees of freedom of the vector bosons $W_\pm$ and $Z$. At the price of loosing manifest gauge invariance, one can deal directly, at the EFT level, with the observable (and now finally all observed) low-energy degrees of freedom of the theory: three massive vector bosons and one Higgs particle.

The use of the unitary gauge brings into cosmology similar advantages. The operators can be organized straightforwardly in powers of the number of perturbations; as a consequence, operators beyond the linear order do not affect the background evolution. Moreover, the terms in the expansion have direct observable implications. For instance,  the cubic operators of the EFT of inflation formalism can be straightforwardly  related to the observable three-point functions of the CMB \cite{EFT2,Senatore:2009gt,Creminelli:2010qf,daniel}.

The application of this formalism to dark energy was initiated in \cite{Creminelli:2008wc}, where generic minimally coupled single-field models were considered. The main difference with inflation is that in the late time universe matter species are present.
In this paper we extend the study of \cite{Creminelli:2008wc} to a non-minimally coupled scalar field in order to develop a unifying theoretical framework aimed to include {\it all} dark energy and modified gravity models. The logic of our approach can be summarized in the following two basic steps:
\begin{itemize}
\item[a)] We assume the validity of the weak equivalence principle (WEP)\footnote{With this choice, we aim to emphasize the ``modified gravity'' content of the theory and disentangle our analysis from the interesting but somewhat off-topic rich phenomenology of WEP violations. The choice of a universally coupled metric is stable under radiative corrections in the matter sector~\cite{lamnic,picon,NP}, meaning that WEP violations are generally expected to be delivered by Planck suppressed operators. See Sec.~\ref{sec_einst} for a relaxation of this assumption.} and therefore the existence of a metric $g_{\mu \nu}$ universally coupled to matter fields $\psi_m$ through an action $S_m[g_{\mu \nu},\psi_m]$; 
\item[b)] We write the  unitary gauge action, i.e.~the most general gravitational action for such a metric compatible with the residual symmetries of unbroken spatial diffeomorphisms.  
\end{itemize}
As emphasized in~\cite{EFT1,EFT2}, the last point allows, although does not postulate from the beginning, the presence of a scalar field $\phi$ in the DE sector. In unitary gauge such a scalar  does not appear explicitly because it is ``eaten'' by the metric.  More practically, the time coordinate  is chosen such that it is a function of the scalar field, so that  the fluctuations of $\phi$ around the background vanish,\footnote{We implicitly assume that $\phi_0(t)$ is a monotonic function of time in the relevant time interval.} $\delta \phi(t, \vec x) \equiv \phi(t, \vec x) - \phi_0(t) =0$.
The scalar degree of freedom can reappear explicitly in the action, together with full diffeomorphism invariance, by performing the ``Stueckelberg trick'', i.e.~by performing an infinitesimal time diffeomorphism $t\rightarrow  t + \pi(x)$ (see Sec.~\eqref{sec_guide}), where $\pi$ is now the field perturbation encoding the scalar dynamics of DE.

In order to show the unifying power of this approach, let us  write the unitary gauge DE action in terms of the universally coupled Jordan metric $g_{\mu \nu}$. This reads 
\begin{equation} \label{example}
S =  \int d^4x \sqrt{-g} \left[\frac{\MM^2}{2} f(t) R - \Lambda(t) - c(t) g^{00}\right] + S_{DE}^{(2)}\; ,
\end{equation}
where $R$ is the Ricci scalar, $f$, $\Lambda$ and $c$ are functions of the time coordinate $t$, $g^{00}$ is the upper time-time component of the Jordan frame metric and $S_{DE}^{(2)}$ indicates terms that start explicitly quadratic in the perturbations and therefore do not affect the background. More explicitly, the last three lines of eq.~\eqref{ac} below. In this paper, $\MM$ is the ``bare'' Planck mass. Its relation with the observed gravitational constant is given in some specific cases in eqs.~\eqref{M-Mp}, \eqref{M-Mp2} and \eqref{M-Mp_total}. 

Let us first consider the square brackets. The time-dependent coefficient $f(t)$ in front of the Ricci scalar is reminiscent of the non-minimal coupling function in scalar-tensor theories, except that here it is written in unitary gauge.
The presence of this function is the main difference with the EFT of quintessence approach \cite{Creminelli:2008wc}.  In the absence of matter fields, as discussed in~\cite{EFT2}, $f(t)$ can be re-absorbed by a conformal transformation of $g_{\mu \nu}$. 
Here, however, the matter sector uniquely singles out the Jordan 
metric $g_{\mu \nu}$, in which test particles follow geodesics.  In this paper we mostly privilege the Jordan frame but we provide the reader with the equivalent, and often more handy, ``Einstein''-frame version of our formalism in Sec.~\ref{sec_einst}. 

The other two functions of time, $\Lambda(t)$ and $c(t)$, have also a very intuitive origin in the simple case of a canonical scalar field. 
For instance, in unitary gauge the kinetic scalar term $(\partial \phi)^2$ becomes
\begin{equation}\label{g00}
-\frac{1}{2} (\partial \phi)^2 \ \equiv \ - \frac{1}{2} g^{\mu \nu} \partial_\mu \phi \partial_\nu \phi \ \rightarrow \ - c(t) g^{00}\, ,
\end{equation}
with $c(t) = \dot \phi_0^2/2$. However, the operator $- c(t) g^{00}$ can also receive contributions by higher-order operators such as $(
\partial \phi)^4$. In general, $\Lambda(t)$ and $c(t)$
can be expressed in terms of background quantities such as the DE background energy density and pressure, respectively  $\rho_D$ and $p_D$, and the Hubble rate $H$.
In Sec.~\ref{sec_general_1}, by using the Einstein equations we find
\be
 \Lambda (t) = \frac12\left[\rho_D -  p_D + \MM^2 (5 H \dot f +  \ddot f)\right] \;, \label{8} \quad
 c (t) = \frac12\left[ \rho_D + p_D + \MM^2 (H \dot f - \ddot f)\right]  \;.
\ee
The background dark energy density and pressure   are defined by the modified Friedmann equations~\eqref{frie1} and~\eqref{frie2} below and are thus derivable from the observed DE equation of state.

Of course, structure formation will be also sensitive  to the quadratic and higher-order terms contained in $S_{DE}^{(2)}$ and these will vary from one DE model to the other. 
In our EFT approach, the building blocks by which 
these operators are constructed are perturbations of quantities which are invariant under spatial diffeomorphisms, such as  the upper time-time component of the metric $g^{00}$, the extrinsic curvature of  uniform-time hypersurfaces  $K^{\mu}_{\ \nu}$ (esplicitely defined in eq.~\eqref{definition_ext}) and the  Riemann tensor $R_{\mu \nu \alpha \beta}$ \cite{EFT2}. We are also allowed to contract these quantities with the metric or among each other to form scalars and to take their derivatives. 
These perturbations appear in quadratic or higher-order combinations so that they
do not affect the background and can be studied---and fitted against observations---independently and in a systematic way.
Despite the richness of $S_{DE}^{(2)}$, 
the higher-order  operators  are classified according to their perturbative effects so that at the linear order in the perturbations only a finite number of them have to be considered. 
We discuss all these higher-order operators in details in Sec.~\ref{sec_secondorder} and below.

In summary, including higher-order terms our Jordan frame action reads 
\begin{equation} 
\begin{split}
S&=  \, \frac12\int d^4x \sqrt{-g} \left[\MM^2  f R \,  - \rho_D +  p_D - \MM^2 (5 H \dot f +  \ddot f)    -  \left(\rho_D + p_D + \MM^2 (H \dot f - \ddot f) \right)  g^{00} \right. \\[2mm]
&+ \,  M_2^4 (\delta g^{00})^2 
  - \bar m_1^3\,  \delta g^{00} \delta K - \bar M_2^2\,  \delta K^2 - \bar M_3^2\,  \delta K_{\mu}^{\ \nu} \delta K_{\ \nu}^\mu + m_2^2 h^{\mu \nu} \partial_\mu g^{00} \partial_\nu g^{00}\\[1mm]
  &\ \ +\lambda_1 \delta R^2 + \lambda_2 \delta R_{\mu \nu} \delta R^{\mu \nu} + \mu_1^2 \delta g^{00} \delta R +\gamma_1  C^{\mu \nu \rho \sigma} C_{\mu \nu \rho \sigma} + \gamma_2  \epsilon^{\mu \nu \rho \sigma} C_{\mu \nu}^{\ \ \ \kappa \lambda} C_{\rho \sigma  \kappa \lambda} \\
  &+  \left. \frac{M_3^4}{3} (\delta g^{00})^3 - \bar m_2^3\,  (\delta g^{00})^2  \delta K  + \dots \, .
  \right]  \;, \label{ac}
\end{split}
\ee
where $\delta g^{00} \equiv g^{00} +1 $, $\delta K^{\mu}_{\ \nu}$ is the perturbation of the extrinsic curvature, $\delta K$ its trace,  $\delta R$ and $\delta R_{\mu \nu} $ the perturbations of the Ricci scalar and tensor, respectively, and $C_{\mu \nu \rho \sigma}$ is the Weyl tensor. $M_i$, $m_i$, $\bar M_i$, $\bar m_i$ and $\mu_i$ are mass parameters while $\lambda_i$  and $\gamma_i$ are dimensionless parameters; they can all depend on time.

Here follow some explanations, comments and a summary of results. 

\begin{list}{\labelitemi}{\leftmargin=1em}
\item \emph{Set up.\ } The action~\eqref{ac} is especially set up and arranged for cosmological perturbations on a FRW background. Every term can be expanded to arbitrary order.
However,  all DE operators contributing up to linear order in the perturbations are contained in the first line of eq.~\eqref{ac} and their coefficients are fixed by background quantities only, $\rho_D$, $p_D$, $H$ and $f$, which can be fitted against observations (Sec.~\ref{sec_general_1}).
Quadratic terms (second and third lines) and higher (forth line, etc.) leave the background unchanged. Their time-dependent coefficients,  $M_2^4$, $\bar m_1^3$ etc., are sensitive to  tests which depend on the clustering or screening properties of DE.

\item \emph{Jordan frame.\ } The action~\eqref{ac} is written in the Jordan frame $g_{\mu \nu}$ that universally couples to the matter fields.  We privilege the Jordan frame for two reasons:  it is more directly related to observations~\cite{NP} and  it is univocally defined by the coupling to matter, once we postulate the WEP. 

\item \emph{The scalar dynamics and mixing.\ } The action~\eqref{ac} is written in unitary gauge and therefore only depends on the metric and its derivatives.   Full diffeomorphism invariance  can be restored by introducing a scalar degree of freedom $\pi$ {\em via} the Stueckelberg trick (see, however, Sec.~\ref{sec_hidden}), after which  any other gauge can be chosen.  At sufficiently high energy, only the kinetic terms are important in the quadratic action. In Ref.~\cite{EFT2} this is called the decoupling limit, because in the simplest cases---as well as in the notable one of Electroweak $SU(2) \times U(1)$ theory---the scalar decouples from the gauge fields. More generally, 
in Jordan frame the scalar can be kinetically mixed with gravity, in which case there is no decoupling limit. In Sec.~\ref{sec_mixing} we study the mixing due to $\dot f$ and $\bar m_1^3$ in Newtonian gauge. We show that one can de-mix $\pi$ and gravity by an appropriate local field redefinition, at the expense of coupling $\pi$ to matter.

\item \emph{Comparison with EFT of inflation.\ } We are considering few more operators than the ones included in~\cite{EFT2} for the inflationary case. 
In the absence of matter fields, some of them ($f$, $\lambda_1$, $\lambda_2$) can be reabsorbed by a field redefinition of the metric tensor. Others (like $m_2^2$ and $\mu_1^2$) were not \emph{explicitly} included and generally implied by ellipsis; we have included them here because we found them essential for some DE models. As mentioned, we also gave more relevance to the operator $\bar m_1^3$ (there called $\bar M_1^3$ and also discussed in \cite{Creminelli:2008wc}), as it is responsible for  kinetic mixing.

\item \emph{Universality.\ } The action~\eqref{ac} describes virtually all DE models based on a single degree of freedom, as we detail in Sec.~\ref{sec:matching} with several examples. What this action \emph{does not} describe are genuinely higher-dimensional regimes of, say, brane scenarios, that cannot be encoded in a 4-d description, massive gravity away from the decoupling limit, theories including vector fields participating to the gravitational sector such as TeVeS (see e.g.~\cite{costas} for references),  as well as modifications of gravity of the \emph{geometrical} type, such as those suggested in~\cite{usep1,usep2}. As discussed in Sec.~\ref{sec_general}, special and different couplings to the dark matter sector (like those needed for models of ``coupled quintessence", e.g.~\cite{Amendola:1999er,Gasperini:2001pc,Comelli:2003cv}) can also be included with minor modifications.

\item \emph{Action principle vs.\ linear equations and regime of applicability.\ } Despite the technical complications related to the gravitational sector (e.g.~having to solve the constraints~\cite{malda}) the formulation in terms of an action is completely general  and therefore preferable to the linearized equations. The regimes of applicability of~\eqref{ac} span from the weakly coupled largest cosmological scales down to the UV cutoff of the EFT. This can be estimated by computing the energy at which  unitarity is violated \cite{EFT2,ArkaniHamed:2005gu}.\footnote{A conservative statement is to say that our action is valid on wavelengths longer than the virialization scale of dark matter.}
\end{list}
\vspace{.3cm}

Other proposals employing an EFT approach to cosmic acceleration have been put forward in the literature, such as those of Refs.~\cite{PZW,BF,bean}. They are based on a covariant description of a canonical scalar field $\phi$ together with higher-order terms in a covariant expansion up to four derivatives, in the spirit of the EFT of inflation \emph{\`a la} Weinberg \cite{wein-eft}.
This approach is certainly a promising starting point in the direction of unifying models with an effective description. Moreover,
the familiar language of (covariant) EFT, with the relative importances of mass scales and operators clearly displayed, can be used here in all its power to explore the  space of DE models. As shown in this references, such a space can be tighten up by perturbative field redefinitions: models that look different from each other may in fact be equivalent EFTs at a closer scrutiny. 

However, we would like to stress the advantages of our approach as compared to a the aforementioned one, or more generally as compared to the use of a covariant description using the language of $\phi$. Most of these points are also well explained in \cite{EFT2,Creminelli:2008wc,Creminelli:2010qf}.

\begin{list}{\labelitemi}{\leftmargin=1em}
\item \emph{Covariance vs background/perturbations splitting.\ } In the $\phi$ language, any further term in the EFT expansion corrects the FRW background. This means that including a new perturbative correction boils down, in practice, to studying a new model altogether \emph{ab initio}: solving new equations for the background, studying perturbations around it in order to seek for specific dynamical effects, etc. This is avoided in our approach that deals directly with the perturbations. Note however that this does not mean, as sometimes stated, that our approach cannot address the background
dynamics: given action \eqref{ac} with a finite number of operators, we can always construct a covariant action in the language of $\phi$ which correctly reproduces the evolution of the background and of the perturbations, as shown in \cite{Creminelli:2008wc} and in Sec.~\ref{sec_covariant}.
 
\item \emph{Stability.\ } The stability and consistency of any model can only be evaluated once one turns to $\phi$ and $g_{\mu \nu}$ fluctuations expanded around a FRW background. This is also the case of theories that are known to have second-order field equations, such as the Horndeski theory \cite{horndeski,Deffayet:2009mn,fab-four}, but whose stability depends on specific signs of the quadratic kinetic Lagrangian of the fluctuations.

\item \emph{Relevance of operators.\ } The relative importance of the various operators can be assessed only after expanding  the action around a background. For instance, there are cases in which covariant operators that are naively higher order in perturbation theory become in fact relevant and should not be treated as perturbations in the standard EFT sense. For example, in the ghost condensate mechanism~\cite{ghost} a vanishing speed of sound is obtained by the balance of the two equally important $(\partial \phi)^4$ and $(\partial \phi)^2$ terms.  There is also a class of higher derivative `galileonic' terms~\cite{NRT,Deffayet:2009wt}  that can be trusted well beyond the naive EFT expectations, as they lead to evolution equations that are not  higher order in time. These cases  are naturally included in our approach.

\item \emph{Standard form of the action.\ } In the language of $\phi$ one can always make a field redefinition $\phi \to \tilde \phi(\phi)$. The resulting action is different but it  describes the same physics. This redundancy may be easy to capture for a simple enough  Lagrangian but it becomes more difficult for a more general one such as, e.g., ${\cal L}(\phi, (\delta \phi)^2, \square \phi)$. The action~\eqref{ac} is in ``standard'' form, in the sense that there is no residual ambiguity due to possible field redefinitions.

\end{list}

\vspace{.2cm}
\section{A very general DE theory} 

In this section, after a brief presentation of the unitary gauge formalism, we introduce our DE action. First we discuss the operators that can be fixed by the background equations of motion and then we discuss higher-order terms. 

\subsection{Quick guide to unitary gauge}
\label{sec_guide}

While referring the reader to~\cite{EFT1,EFT2} for more details, it is useful here to give an idea of how the unitary gauge works. In a general (perturbed) FRW universe, $\phi(t, \vec x) = \phi_0(t) + \delta \phi(t, \vec x)$. By choosing the coordinate $t$ to be a function of $\phi$, $t = t(\phi)$, we thus simply have $\delta \phi = 0$. Therefore, as mentioned, the DE action written in this gauge only displays metric degrees of freedom. For instance, in eq.~\eqref{g00} we have already seen how
the kinetic scalar term $(\partial \phi)^2$ is transformed in unitary gauge.

For concreteness, we have made touch with the well known example of a scalar field. But what $\phi$ really does for us here is to define a preferred ($\phi = const.$) time slicing. One can well forget about $\phi$ and build a Lagrangian with the unit vector $n_\mu$ perpendicular to such a slicing.  In unitary gauge,
\begin{equation} \label{nmu} 
n_\mu \ \equiv \ - \frac{\partial_\mu \phi}{\sqrt{- (\partial \phi^2)}} \ \rightarrow \ - \frac{\delta_\mu^0}{\sqrt{-g^{00}}}\; .
\end{equation}
It follows that in the DE action, beside genuinely 4-d covariant terms such as the Ricci scalar $R$, also contractions of tensors with $n_\mu$ are allowed, i.e., by~\eqref{nmu}, tensors with free upper 0 indices such as $g^{00}$, $R^{00}$, etc. Moreover, because time-translations are broken, the coefficients of the operators in our action are allowed to be time dependent. These two remarks are both exemplified by~\eqref{g00}. 
Covariant derivatives of $n_\mu$ can also be used to express operators in the DE action. Equivalently, we can use their projection along and orthogonal to $t = const$ surfaces, i.e.~the extrinsic curvature
\begin{equation} \label{definition_ext}
K_{\mu \nu} \equiv h^{\ \sigma}_\mu \nabla_\sigma n_\nu\; ,
\end{equation}
where $h_{\mu \nu} \equiv g_{\mu \nu} + n_\mu n_\nu$ is the induced spatial metric, and  $n^\sigma \nabla_\sigma n_\nu \propto h^{\ \mu}_\nu \partial_\mu g^{00}$.

Diffeomorphism invariance is restored by the ``Stueckelberg trick'', i.e.~by performing an infinitesimal time diffeomorphism $t\rightarrow  t+\pi(x)$, $\vec x \rightarrow \vec x$. After a trivial coordinate redefinition one finds, as a new field in the action,  $\pi(x)$. This  makes apparent the presence of a scalar degree of freedom in the DE sector. From the action written in unitary gauge, terms containing $\pi$ are generated by the Stueckelberg trick in two ways. First, any time-dependent coefficient, such as $c(t)$ in~\eqref{g00}, generates terms in which $\pi$ appears not derivated:
\begin{equation}
c(t) \ \rightarrow \ c(t + \pi) \ = \ c(t) + \dot c(t)\,  \pi + \frac12 \ddot c(t) \, \pi^2 + \dots \;.
\end{equation}
Second, operators that are not 4-d diffeomorphism invariant transform under time diffeomorphisms and thus generate terms with derivatives acting on $\pi$. Few relevant examples are:
\begin{align}
g^{00} \ &\rightarrow \ g^{00} + 2 g^{0 \mu} \partial_\mu \pi + g^{\mu \nu} \partial_\mu \pi \partial_\nu \pi\; , \label{stuck1} \\  
g^{0i} \ &\rightarrow \ g^{0i} + g^{\mu i} \partial_\mu \pi \;,\\
\delta K_{ij} \ &\rightarrow \ \delta K_{ij}  - \dot H  \pi h_{i j}   - \partial_i \partial_j  \pi\;, \\
\delta K\  &\rightarrow \ \delta K - 3 \dot H \pi - a^{-2} \nabla^2 \pi \;, \label{stuck4}
\end{align}
where in the last two lines we have expanded at  linear order in $\pi$.

\subsection{Structure of the DE action: background equations} \label{sec_general_1}

Once we assume that there is a metric $g_{\mu \nu}$ minimally coupled to matter fields, we can proceed and build the most general unitary gauge action $S[g_{\mu \nu};t]$ to describe the dark energy-modified gravity sector. In doing this, we can follow step by step the construction outlined in~\cite{EFT2}, and the general rules there given; with two exceptions:
\begin{enumerate}
\item We now need to allow a general free function of time $f(t)$ in front of the Ricci scalar. In fact, we cannot do any field redefinition of the metric tensor if we want to stick with the Jordan frame metric that minimally couples to matter.
\item For the same reason above,  in~\eqref{ac} we are not allowed perturbative field redefinitions of the metric that re-absorb quadratic curvature terms~\cite{cliff} such as $\lambda_1$ and $\lambda_2$.
\end{enumerate}
This procedure univocally determines the only three operators that contain up to linear order terms in the perturbations. Schematically, 
\begin{equation} \label{example2}
 S =  \int d^4x \sqrt{-g} \left[\frac{\MM^2}{2} f(t) R - \Lambda(t) - c(t) g^{00}\right] + S_{DE}^{(2)}\; .
\end{equation}
The universality of the above action is also proved independently by going through the Einstein frame construction of Sec.~\ref{sec_general} and then transforming it back to the Jordan frame (Sec.~\ref{sec_etoj}).
In unitary gauge $S$ only depends on the metric, which means that we are implicitly allowing, beside Einstein gravity, up to one more scalar degree of freedom in the DE sector.

Action~\eqref{example2} closely resembles a Brans-Dicke theory~\cite{BD,polarski}; it is straightforward to recognize 
the meaning of our coefficients $f$, $c$ and $\Lambda$ once we take a Brans-Dicke theory and go to unitary gauge. Perhaps slightly less intuitive is 
to see how more ``exotic'' operators fit into our scheme. While devoting Sec.~\ref{sec:matching} to translate several DE models in our language, here it is worth 
considering, just as an example, the case of an operator for which the correspondence with~\eqref{example2} is less obvious. A natural generalization of the Brans-Dicke coupling $\phi R$, is $X R$, where $X \equiv (\partial \phi)^2$. 
Because in unitary gauge $X = \dot \phi_0^2 g^{00}$, and $g^{00} = -1 + \delta g^{00}$, the operator in question is expressed in unitary gauge as 
\begin{equation}
X R\ \ = \ \dot \phi_0^2 \left[-R + R^{(0)}(t) + R^{(0)}(t) g^{00} + \delta g^{00} \delta R\right],
\end{equation}
with $R^{(0)}$ the background value of the Ricci scalar. From the above expression, it is immediate to recognize the contributions of $XR$ to $f$, $c$ and $\Lambda$ in equation~\eqref{example2}. The last term in square brackets is quadratic in the perturbations and contribute to the coefficient $\mu_1^2$ in eq.~\eqref{ac}.

It is indeed remarkable that, up to linear order in perturbations, arbitrary complicated covariant operators only contribute to the square bracket of eq.~\eqref{example2}. 
In what follows we show how to fix the coefficients $c$ and $\Lambda$ above by using the background equations of motion. In the next subsection we discuss quadratic and higher-order operators.

Let us start fixing $c$ and $\Lambda$ by using the equations for a general FRW background of constant spatial curvature $k$. 
In order to do so, we can neglect quadratic and higher-order terms contained in $S_{DE}^{(2)}$. By varying with respect to $g^{\mu \nu}$ we obtain
\be \label{einsteine}
\left( f G_{\mu \nu} - \nabla_\mu \nabla_\nu f + g_{\mu \nu} \square f \right)  \MM^2 + ( c g^{00} +  \Lambda) g_{\mu \nu} -2 c  \delta^0_\mu \delta^0_\nu = T_{\mu \nu}\;,
\ee
where $T_{\mu \nu}$ is the energy-momentum tensor of matter. We can take, for instance, the trace and the $00$ component of the above equation and consider that, on the background, 
\begin{equation} \label{backvalues}
G_{00} = 3 H^2 + 3 \frac{k}{a^2},\quad R = -G_\mu^\mu = 12 H^2 + 6 \dot H + 6 \frac{k}{a^2}, \quad \square f(t) = -\ddot f - 3 H \dot f. 
\end{equation}
By making the perfect-fluid approximation for matter fields, $T_\mu^\nu = {\rm diag}(-\rho_m, 
p_m, p_m, p_m)$, we finally obtain,  
\begin{align}
c &= \MM^2 f \left( -  \dot H + \frac{k}{a^2} -  \frac12 \frac{\ddot f}{f} +  \frac{H}{2} \frac{\dot f}{ f} \right)  - \frac12 (\rho_m+p_m)  \;, \\
\Lambda &= \MM^2 f \left(\dot H +3 H^2 +2 \frac{k}{a^2} + \frac12 \frac{\ddot f}{f}+\frac{5 H}{2}   \frac{\dot f}{f} \right)  - \frac12 (\rho_m-p_m)  \;.
\end{align}

As already mentioned, the advantage of dealing with Jordan frame quantities is a more direct connection with observations. 
For instance, the Hubble constant $H_0 = 72 \, \pm \, 8 \, {\rm km} / {\rm s}\, {\rm Mpc}^{-1}$ measured by the Hubble Space Telescope~\cite{h0} translates into a statement about the \emph{Jordan} metric (Hubble parameters in Einstein and Jordan frames are related by eq.~\eqref{hubbles}). We also refer the reader to the discussion in~\cite{NP}. In Jordan frame, 
test particles follow geodesics and the energy-momentum tensor in the matter sector is conserved in the usual way, 
\be
\label{cons_m}
\dot \rho_m + 3 H (\rho_m + p_m) =  0\; .
\ee

In contrast, in presence of a coupling to gravity there is no conserved DE density $\rho_D$ and pressure $p_D$. However, it proves useful to \emph{define}\footnote{Equivalently, one could define $\rho_D^{\rm eff}$ and $p_D^{\rm eff}$ as in~\cite{post-fried1,post-fried2}, such that
\begin{equation}
\rho_D = f\rho_D^{\rm eff}+ (f-1) \rho_m  , \qquad p_D = f p_D^{\rm eff} + (f-1) p_m \, .
\end{equation}
In this case eqs.~(\ref{frie1})--(\ref{frie2}) get closer to the standard form,
\be
 H^2 + \frac{k}{a^2}    =  \frac1{3 \MM^2} (\rho_m + \rho^{\rm eff}_{D}  )  \; ,\qquad
\dot H -  \frac{k}{a^2} =  - \frac1{2  \MM^2} (\rho_m + \rho^{\rm eff}_{D} +p_m + p^{\rm eff}_{D}  )  \;.
\ee} the latter through the following (modified) Friedmann equations
\begin{align}
 H^2 + \frac{k}{a^2}    &=  \frac1{3 f \MM^2} (\rho_m + \rho_{D}  )  \label{frie1}\; ,\\
\dot H -  \frac{k}{a^2} &=  - \frac1{2 f \MM^2} (\rho_m + \rho_{D} +p_m + p_{D}  )  \label{frie2}\;.
\end{align}
Indeed, by using~\eqref{einsteine} and~\eqref{cons_m}, we obtain the modified conservation equation
\begin{equation} \label{conservation}
\dot \rho_D + 3 H (\rho_D + p_D) = 3 \MM^2 \dot f  \left(H^2 + \frac{k}{a^2}\right)\, . 
\end{equation}
Finally, using~\eqref{frie1}-\eqref{frie2}, we find 
\begin{align}
c \ &  =   \   \frac12 (  -  \ddot f  +  H \dot f ) \MM^2  + \frac12 (\rho_{D}+p_{D}) \;, \label{c2}\\
\Lambda \, &  = \ \frac12 (   \ddot f  + 5 H \dot f ) \MM^2 + \frac12 (\rho_{D}- p_{D})     \;. \label{L2}
\end{align}
The above are used to obtain the first two lines of eq.~\eqref{ac}.
The only freedom left in fixing $c$ and $\Lambda$ is thus the function $f(t)$. In practice, however, what discriminates among the different DE models is the choice of few parameters of order one. For instance, by expanding in the proper time around the present epoch, 
\begin{equation}
f(t) = 1 + f_1 H_0 (t-t_0) + f_2 H_0^2 (t-t_0)^2 + \dots\; .
\end{equation}
In turn, the parameters $f_i$ can be constrained by post-Newtonian tests of gravity (see Sec.~\ref{sec_fdot}). Clearly, an analogous expansion should apply to all time dependent coefficients appearing in our action.

\subsection{Quadratic and higher-order operators}
\label{sec_secondorder}

The non-universal part of the DE action, $S_{DE}^{(2)}$, is at least quadratic in the perturbations:
\be
S_{DE}^{(2)} =  \int d^4x \sqrt{-g} \, F^{(2)}  (\delta g^{00}, \delta K_{\mu \nu}, \delta R_{\mu \nu \rho \sigma}; h^{\ \mu}_{\nu} \nabla_\mu ; t )\;, 
\ee
where $F^{(2)}$ is a polynomial of at least quadratic order in the arguments. 
The first argument is $\delta g^{00} \equiv g^{00}+1$; hence, $F^{(2)}$ includes terms such as
\be
F^{(2)} \, \supset \, \frac{M_2^4}{2} (\delta g^{00})^2 +  \frac{M_3^4}{3!} (\delta g^{00})^3 + \, \ldots \;,
\ee
where $M_2^4$, $M_3^4$, etc., are functions of time. In this formalism, a $P(\phi,X)$ theory looks like an expansion in $\delta g^{00}$ as above (see Sec.~\ref{sec_quint}).

By considering  derivatives projected on the surfaces of constant $t$, $F^{(2)}$ will include also terms such as
\be
F^{(2)} \, \supset \, \frac{ m_{2}^2}{2} h^{\mu \nu} \partial_\mu g^{00} \partial_\nu g^{00} + \, \ldots\;.
\ee
Although such terms have not been particularly discussed in the original EFT of inflation papers, they are of relevance, for instance, for Lorentz violating models of DE (see Sec.~\ref{sec_khro}). 

The second argument of $F^{(2)}$, $\delta K_{\mu \nu} = K_{\mu \nu} - K^{(0)}_{\mu \nu}$ ($K^{(0)}_{\mu \nu} = H h_{\mu \nu}$), is the perturbation of the extrinsic curvature of surfaces of constant $\phi$, defined in eq.~\eqref{definition_ext}. We show in App.~\ref{sec_invariants} how such tensorial perturbations are defined here. With the addition of this operator and its contractions we generate also terms such as 
\be
F^{(2)} \, \supset \, - \frac{\bar m_1^3}{2} \delta g^{00} \delta K - \frac{\bar M_2^2}{2} \delta K^2 - \frac{\bar M_3^2}{2} \delta K_{\mu}^{\ \nu} \delta K_{\ \nu}^\mu + \ldots\;.
\ee
The operators $\bar M_2^2$ and $\bar M_3^2$ are those appearing in the ghost condensate \cite{ghost}. We will discuss $\bar m_1^3$, which appear in `galileon'  and braiding models, in Sec.~\ref{sec_gal}.

It is worth asking why a term that starts linear in the perturbations such as $K$ \emph{does not} appear in the universal part of~\eqref{example2}. Following  appendix B of Ref.~\cite{EFT2} we can indeed rewrite it as
\begin{equation} \label{80A}
\int \! d^4 x \sqrt{-g} \, l(t) K \equiv \int \! d^4 x \sqrt{-g}\, l(t) \nabla_\mu n^\mu = - \int \! d^4 x \sqrt{-g} \; n^\mu \, \partial_\mu l = - \int \! d^4 x \sqrt{-g} \sqrt{-g^{00}} \dot l\, .
\end{equation}
This relation will be often used in the following of the paper. Note that there is a sign difference with respect to eq.~(80) of Ref.~\cite{EFT2}, due to the different convention for the sign of $n_\mu$ defined in eq.~\eqref{nmu} above.

Finally, also 4-d curvature invariants are allowed. These are gauge invariant terms and, as such, do not produce any $\pi$ through the Stueckelberg trick.  In the absence of matter, it is always possible to re-absorb the terms quadratic in the curvature with a perturbative field redefinition of $g_{\mu \nu}$ (see e.g.~\cite{cliff}). However, in the presence of matter this would generate scalar couplings to matter fields, which would make us exit the Jordan frame. In App.~\ref{sec_invariants} we show how to express curvature invariants in terms of the zero and first-order terms and in terms of quadratic operators such as
\be
F^{(2)} \, \supset \, \lambda_1 \delta R^2 + \lambda_2 \delta R_{\mu \nu} \delta R^{\mu \nu} +\gamma_1  C^{\mu \nu \rho \sigma} C_{\mu \nu \rho \sigma} + \gamma_2  \epsilon^{\mu \nu \rho \sigma} C_{\mu \nu}^{\ \ \ \kappa \lambda} C_{\rho \sigma  \kappa \lambda} + \dots\;,
\ee
where instead of  the perturbation of the Riemann tensor $\delta R_{\mu \nu \rho \sigma}$ we are using the Weyl tensor $C_{\mu \nu \rho \sigma}$ \cite{wein-eft}, defined by
\be
C_{\mu \nu \rho \sigma} \equiv R_{\mu \nu \rho \sigma} - \frac12 (g_{\mu \rho} R_{\nu \sigma} - g_{\mu \sigma} R_{\nu \rho} - g_{\nu \rho} R_{\mu \sigma} + g_{\nu \sigma} R_{\mu \rho}) + \frac{R}{6} (g_{\mu \rho} g_{\nu \sigma} - g_{\nu \rho} g_{\mu \sigma})\;. \label{Weyl}
\ee
This vanishes on a FRW background and thus do not need to be perturbed. Therefore, all these terms  are also included in our action.

\subsection{Comparison with the covariant approach}
\label{sec_covariant}

As already observed, one of the advantages of our EFT is that only three functions specifying the action \eqref{ac}, i.e.~$f(t)$, $\Lambda(t)$ and $c(t)$, are completely fixed by the background evolution. All other operators leave the background unchanged and can only be constrained by observations involving metric and $\pi$ fluctuations, at an increasingly high order in perturbation theory. For instance, only the operators in the second and third lines of eq.~\eqref{ac} affect the linear  evolution of perturbations, the other functions entering only at second or higher order.

This is different from the standard descriptions of DE as a scalar field $\phi$ in terms of a covariant Lagrangian.  For instance in $k$-essence one usually deals with the whole function $P(\phi, (\partial \phi)^2)$. Although it is elegant to have a covariant description potentially encompassing a variety of solutions,
an EFT approach should just content itself with the less ambitious task of providing a theoretical framework which correctly reproduces cosmological observations, given a few parameters.

Effective field theory can also be used directly at the level of the fully covariant theory.  In cosmology, such an approach has been advocated and developed by  Weinberg in the context of inflation~\cite{wein-eft} and soon after extended to DE  by Park, Zurek and Watson~\cite{PZW} and Bloomfield and Flanagan~\cite{BF}, in Jordan and Einstein frames respectively. Again, since most DE models reduce to a scalar tensor theory in their relevant regimes, one writes, schematically
\begin{equation} \label{covariant}
S=   \int d^4x \sqrt{-g} \left[\frac{\MM^2}{2} f(\phi) R -\frac12  (\partial \phi)^2 - V(\phi)+ {\cal F}[\phi,g^{\mu\nu}]\right]  \, .
\end{equation}
What characterizes different models is the specific choice of $f(\phi)$, $V(\phi)$ and ${\cal F} [\phi, g^{\mu\nu} ]$, a local scalar function of $\phi$, $g_{\mu \nu}$ and their derivatives.

Here we show that, given our unitary gauge action \eqref{ac} with a finite order of operators, one can construct a covariant EFT action which correctly reproduces the unperturbed and perturbed evolution at that given order (see also \cite{EFT2,Creminelli:2008wc}). Being covariant, this description unifies the evolution of the background with that of perturbations.

Let us consider eq.~\eqref{ac} up to second order in our expansion, i.e.~the first, second and third lines of this equation and no other higher-order operator. To make this action covariant we perform the following replacements:
\be
\begin{split}
t  \to & \ \frac{\phi}{\MM^2}\;, \\
\delta g^{00}  \to & \ 1+ \frac{(\partial \phi)^2}{ \MM^{4}}\; \\
\delta K_\mu^{\ \nu}  \to & \ -  H (\phi) h_\mu^{\ \nu} - \frac3{2 \MM^2} \nabla_\mu \partial^\nu \phi \;,\\
\delta K  \to & \ - 3 H (\phi) - \frac3{2 \MM^2} \square \phi - \frac1{2 \MM^6} \left[ (\partial \phi)^2 \square \phi + \partial^\mu \phi \partial_\mu (\partial \phi)^2 \right]\;, \\\delta R  \to & \ R - R^{(0)}\;, \qquad \delta R_{\mu \nu}  \to R_{\mu \nu} - R_{\mu \nu}^{(0)}\;,
\end{split}
\ee
where $R^{(0)}$ and $R_{\mu \nu}^{(0)}$ are the background values of $R$ and $R_{\mu \nu}$, respectively given in eqs.~\eqref{R0} and \eqref{R0munu} of App.~\ref{sec_invariants}.
The terms in eq.~\eqref{ac} containing derivatives of $g^{00}$ only contribute to  order higher than fourth in the covariant derivative expansion, and can thus be dropped.  
Keeping only terms up to fourth order in derivatives and integrating by parts several times, it is lengthy  but straightforward to derive the following covariant action written in Jordan frame,
\be
\begin{split}
\label{ac_cov}
S = & \ \frac12 \int d^4 x \sqrt{-g} \Big[ \MM^2f(\phi) R -  Z (\phi) (\partial \phi)^2 -2V(\phi) \\ & +  a_1 (\phi)(\partial \phi)^4 + a_2 (\phi) (\partial \phi)^2 \square \phi + a_3(\phi) (\square \phi)^2  \\ 
& + b_1 (\phi) R^2 + b_2 (\phi) R_{\mu \nu} R^{\mu \nu}  + b_3 (\phi) R (\partial \phi)^2 \\
& +c_1(\phi) C^{\mu \nu \rho \sigma} C_{\mu \nu \rho \sigma} + c_2 (\phi) \epsilon^{\mu \nu \rho \sigma} C_{\mu \nu}^{\ \ \ \kappa \lambda} C_{\rho \sigma  \kappa \lambda} \Big] \;.
\end{split}
\ee
By an appropriate choice of functions $a_i(\phi)$, $b_i(\phi)$ and $c_i(\phi)$,\footnote{The functions $a_i$ and $b_i$  in the covariant action \eqref{ac_cov} are related to the parameters of action \eqref{ac} by 
\be
\begin{split}
f  = &\ 1 -  \frac{12 \lambda_1 (2 H^2 + \dot H) + 2 \lambda_2 (3 H^2 + \dot H) - \mu_1^2 }{\MM^2}  \;,\\ 
-2 V  = & \ - 2 \Lambda + M_2^4 + 3 H \bar m_1^3 - 3 H^2 (3 \bar M_2^2+ \bar M_3^2) + 36 \lambda_1 (2 H^2 + \dot H)^2 \\& \ + 12 \lambda_2 (3 H^4 + \dot H^2 + 3 H^2 \dot H) -6 \mu^2_1 (2 H^2 + \dot H) \;, \\
- Z  = & \ - \frac{2 c - 2 M_2^4 - 3 H \bar m_1^3 +6 \mu^2_1 (2 H^2 + \dot H)}{\MM^4} - \frac{ \bar m_1^3{}' - 9 ( H \bar M_2^2)' - 3(H \bar M_3^2)'}{\MM^2} + 4 (\lambda_2 \dot H)'' \;,\\
a_1  = & \  \frac{ M_2^4}{{\MM^8}} +  \frac{ 3(H \bar M_2^2)' + (H \bar M_3^2)' - (1/2) (\bar M_3^2)'' }{\MM^6} \;,\quad 
a_2  =   \ \frac{\bar m_1^3 }{\MM^6} -\frac{3 (\bar M_3^2)'/2 +4 (\lambda_2 \dot H)'}{\MM^{4}}  \;, \quad a_3  = \ -  \frac{ \bar M_2^2+\bar M_{3}^{2}}{ \MM^4}  \;, \\ 
b_1  = & \ \lambda_1 \;, \quad b_2 = \lambda_2\;, \quad b_3 = \frac{\mu_1^2}{\MM^4}\;, \quad c_1 = \gamma_1 \;, \quad c_2 = \gamma_2\;,
\end{split}
\ee
where a prime denotes the derivative with respect to $\phi$ and all the quantities on the right hand side are now functions of $\phi$ through the substitution $t\to \phi \to \frac{\phi}{\MM^2}$ and a flat FRW background is assumed.} this action exactly reproduces the  background and linear dynamics of \eqref{ac}.
Moreover, by a number of field-redefinitions it can  be reduced to the final covariant EFT actions obtained in \cite{PZW,BF}.

\section{Scalar dynamics and mixing with gravity}
\label{sec_mixing}

As already noted, by writing our most general DE action in unitary gauge, we are allowing the presence of an additional scalar degree of freedom in the DE sector. In most cases, this can be explicitly seen  by applying the Stueckelberg trick, i.e., by forcing a time diffeomorphism 
$t \rightarrow t + \pi(x)$ upon the unitary gauge action~\eqref{ac}, as outlined in Sec.~\ref{sec_guide}.

The simplest way to generate a dynamical $\pi$ field is considering in our action \eqref{ac} a non-vanishing $c(t)$. In this case the Stueckelberg trick generates $\pi$ with a relativistic kinetic Lagrangian $\dot \pi^2 - (\vec \nabla \pi)^2$. Another simple example  is adding the operator $M_2^4$. We will consider a linearly perturbed flat FRW metric  in Newtonian gauge, focussing on scalar perturbations only, in which case
\begin{equation} \label{newtonian} 
ds^2 = -(1+2\F)dt^2 + a^2(t) (1-2 \PP) \delta_{i j} dx^i dx^j\, .
\end{equation}
One way of studying the mixing is considering the linearized equations, which we derive in App.~\ref{app_linear}. In this section we work directly at the level of the quadratic action. 
By making use of eq.~\eqref{stuck1} and of the expression for $g^{00}$, one finds $\delta g^{00} \to 2(\Phi -\dot \pi) + 4 \Phi \dot \pi - \dot \pi^2 + a^{-2}(\vec  \nabla \pi)^2$. Thus, the Lagrangian of $\pi$ reads, neglecting the expansion of the Universe,
\be
-c \; \delta g^{00} + \frac{M_2^4}{2} (\delta g^{00})^2 \overset{\rm kinetic}{\to} (c + 2 M_2^4) \dot \pi^2 -  c ( \vec \nabla \pi)^2   - 4(c+ M_2^4) \dot \pi \Phi\;.
\ee  
At high energy the last term can be neglected, $\pi$ decouples from gravity and propagates with a speed of sound $c_s^2 = c/(c + 2 M_2^4)$. This is the so-called decoupling limit \cite{EFT2}, which takes place at an energy higher than $E_{\rm mix} \sim (c + M_2^4)/[(c + 2 M_2^4)^{1/2} \MM]$. 

While decoupling happens in simple cases such as the one above, in this section we will study examples where the decoupling is not at work and $\pi$ and gravity are mixed already at the kinetic level. In particular, we will study the mixing produced by {\em i}) a non-vanishing $\dot f$; {\em ii}) the operator $\bar m_1^3$. These are operators that produce mixing but do not involve higher derivatives. We will not consider quadratic operators involving higher  {\em spatial} derivatives such as $\bar M_2$ and $\bar M_3$, which in their decoupling limit have been extensively studied in the context of the ghost condensate in \cite{ghost,EFT1,EFT2}, nor $m_2^2$, studied in \cite{Horava:2009uw,Blas:2009qj,Blas:2010hb}. 

Before doing that, let us first mention a well-known case where $\pi$ is not produced {\em via} the Stueckelberg trick and yet the theory does contain a propagating degree of freedom.

\subsection{The ``hidden" scalar}\label{sec_hidden}

By looking at~\eqref{example2}, it is clear that a way~\emph{not to} generate any $\pi$ is having $f$ and $\Lambda$ constant and $c = S_{DE}^{(2)} = 0$, which is the case of $\Lambda$CDM. 
It is useful to briefly review (following e.g.~\cite{ghost,Creminelli:2008wc}) why pure Einstein gravity does not contain any scalar  propagating degree of freedom. 
Let us expand the Einstein-Hilbert action (see App.~\ref{app_formulas}) and retain only the kinetic terms; more specifically, we drop those terms that are subdominant on scales much smaller than Hubble. For $R$, this leaves us  with only the terms displayed in the first line of eq.~\eqref{EH} and we obtain (in this section $\vec \nabla$ is a 3-d gradient and $\nabla^2$ the Laplacian), 
\begin{equation}
 \frac{1}{2}  \sqrt{-g}  R \   \overset{\rm kinetic}{=}   - 3 \dot \Psi^2 +   ( \vec \nabla  \Psi)^2 - 2 \vec \nabla \Phi \vec \nabla \Psi  \, . \label{normal_gravity}
 \end{equation}
By going to Fourier space and diagonalizing the kinetic matrix corresponding to the above expression, we simply obtain the dispersion relation $k^4 = 0$, which means that $\Phi$ and $\Psi$ are non-propagating degrees of freedom.

However, the fact that no $\pi$ is produced by the Stueckelberg trick does not guarantee, by itself, the absence of scalar degrees of freedom. One straightforward example is $F(R)$ gravity. By Taylor expanding around a background value $R^{(0)}$ its Lagrangian reads
\be
\begin{split}
 \frac{1}{2}  \sqrt{-g}  F(R) &=  \frac{1}{2}  \sqrt{-g} (F + F' \delta R + \frac12 F'' \delta R^2 + \ldots ) \\
 &= \frac{1}{2}  \sqrt{-g} (F' R + F - F' R^{(0)}  + \frac12 F'' \delta R^2 + \ldots )\;,
\label{expansionfR}
\end{split}
\ee
where $F'$, $F''$, etc., are computed at $R=R^{(0)}$. The above equation can be reproduced by our action \eqref{example2} with $f=F'$,  $\Lambda = \MM^2(- F + F' R^{(0)})$, $c=0$, $\lambda_1 = \MM^2 F''/2 $, etc.
Since $R$ is a curvature invariant, it does not produce any $\pi$ field under a time diffeomorphism, nor does any function $F(R)$ thereof. 

However, by rewriting the first line of eq.~\eqref{expansionfR} at quadratic order in the perturbations using eq.~\eqref{Ricci_New} in App.~\ref{app_formulas}  and taking the high-energy limit, we find
\begin{equation} \label{eq36}
\frac{1}{2}\sqrt{-g}  F(R) \   \overset{\rm kinetic}{=} \  F' \left[- 3 \dot \Psi^2 +  (\vec \nabla \Psi)^2 -  2 \vec \nabla \Phi \vec \nabla \Psi\right] + 
 F'' \left[3  \ddot \Psi - 2 \nabla^2 \Psi + \nabla^2 \Phi \right]^2 \, .
 \end{equation}
Calculating the determinant of the  Lagrangian in Fourier space one finds 
\be
{\rm det}\; {\cal L} = k^4 F' \left[ 3 F '' (\omega^2  -  k^2 ) - F'  \right]\;.
\ee
Setting this determinant to zero gives the dispersion relation of a relativistic degree of freedom with mass $m^2 = F'/3 F''$.

More explicitly, we can consider the constraint equation $\delta S/\delta \Phi =0$ derived from eq.~\eqref{eq36}. Every term of this equation is acted upon by at least one Laplacian $\nabla^2$ that can be safely integrated, giving
\begin{equation}
\nabla^2 \Phi = 2 \nabla^2 \Psi - 3 \ddot \Psi - \frac{F'}{F''} \Psi\, .
\end{equation}
Plugging this back into the action, integrating by parts  we finally obtain
\begin{equation}\label{psiii}
\frac{\sqrt{-g}}{2} F(R) \   \overset{\rm quad}{=} 3 F' \left[\dot \Psi^2 - (\vec \nabla \Psi)^2\right] - \frac{(F')^2}{ F''} \Psi^2 ,
\end{equation}
the action of a relativistic scalar field. The condition to get to~\eqref{psiii} is that $F''\neq 0$. In Sec.~\ref{sec_f(r)} we outline an equivalent way to deal with $F(R)$ theories which is closer in spirit to our formalism and brings the action to the universal form~\eqref{example2} with $S_{DE}^{(2)} = 0$.

\subsection{Mixing proportional to $\dot f$}
\label{sec_fdot}

We now turn to the situation in which a $\pi$ field is produced \emph{via} Stueckelberg trick and it is kinetically mixed with gravity. Here we consider the well-known case of the mixing produced by a non-constant $f(t)$.  
This is the type of mixing common to all scalar-tensor theories and we just review it here briefly in our language. 
By using the results of App.~\ref{app_formulas}, we write the relevant kinetic terms in Newtonian gauge,
\begin{equation} \label{minklimit}
\begin{split}
S & = \frac{1}{2} \int \sqrt{-g}  \MM^2 f (t) R  \\
 & \!\!\!\!\! \overset{\rm kinetic}{\to }   \int  \MM^2 f \left[ - 3 \dot \Psi^2  - 2 \vec \nabla \F \vec \nabla \PP + (\vec \nabla \PP)^2 + 3 ({\dot f}/{f}) \dot \Psi \dot \pi - ({\dot f}/{f}) \pi (\nabla^2 \F - 2 \nabla^2 \PP)\right] \; .
 \end{split}
\end{equation}
If we go to Fourier space and calculate the determinant of this $3 \times 3$ kinetic matrix we find $\sim k^4 (\omega^2 - k^2)$, signaling the presence of a single propagating degree of freedom with relativistic dispersion relation $\omega^2 = k^2$. 

The transformation that de-mixes $\pi$ from gravity is well known, it is the conformal transformation  to the Einstein frame metric, see Sec.~\ref{sec_einst}. On the metric potentials $\Phi$ and $\Psi$ this transformation reads
\begin{equation}
\label{conf_met}
\Phi_E = \Phi + \frac12 ({\dot f}/{f}) \pi\,, \qquad  \Psi_E = \Psi - \frac12 ({\dot f}/{f}) \pi \;,
\end{equation}
and action \eqref{minklimit} becomes
\begin{equation} \label{mixxing}
\int  \MM^2 f \left[ - 3 \dot \Psi_E^2  - 2 \vec \nabla \F_E \vec \nabla \PP_E + (\vec \nabla \PP_E)^2 + \frac34 ({\dot f}/{f})^2 \left(\dot \pi^2 - (\vec \nabla \pi )^2 \right) \right]  \;.
\end{equation}

Even though in this frame $\pi$ is kinetically decoupled from gravity, it is now directly coupled to matter. 
In order to study the effects  of this theory on matter we consider its Newtonian limit: we can neglect time derivatives and introduce pressureless matter sources.\footnote{We discuss the Newtonian limit in the quadratic approximation. This means that we are neglecting the possible screening effects due to the higher-order operators. } We will do that directly in Jordan frame because this is the frame where test particles follow geodesics.  
For generality, we also add to eq.~\eqref{minklimit} an explicit $\pi$ kinetic term $- c (\partial \pi)^2$. Varying this action with respect to $\Psi$ gives
\be
\label{psiphi}
\Psi = \Phi + ({\dot f}/{f}) \pi\;.
\ee
Plugging this back into \eqref{minklimit} finally yields
\begin{equation}
S = \int  \MM^2 f \left[ - (\vec \nabla \F )^2 - (\dot f/f) \, \vec \nabla \F \vec \nabla \pi -\left((\dot f/f)^2 +  \frac{c}{\MM^2 f} \right) (\vec \nabla \pi)^2 \right] - \F \delta \rho_m\, ,
\end{equation}
where we have added the gravitational coupling to non-relativistic matter $(1/2) T^{00} \delta g_{00}$. 

Varying this action with respect to $\pi$ fixes
\begin{equation} \label{bah}
\pi =  \frac{{- \MM^2 \dot f}}{2 ( c + \MM^2 {\dot f}^2/f  )} \F\, ,
\end{equation}
which can be used in
eq.~\eqref{psiphi} to derive the post-Newtonian parameter $\gamma = \Psi/\Phi$. From~\eqref{psiphi} and~\eqref{bah} we get
\begin{equation}
1 - \gamma =  \frac{\MM^2 {\dot f}^2/f}{2 ( c +  \MM^2 {\dot f}^2/f )} \, .
\end{equation}
By varying with respect to $\F$ we see that the usual Poisson equation for the gravitational potential is modified by a term in $\pi$.
Finally, using \eqref{bah} the Poisson equation reads
\begin{equation}
\bigg(1 - \frac{ \MM^2 {\dot f}^2/f}{4 ( c + \MM^2 {\dot f}^2/f )} \bigg) \nabla^2 \F = \frac{1}{2 \MM^2 f} \delta \rho_m\, .
\end{equation}
By comparison with $\nabla^2 \Phi = \rho_m / 2 M_{\rm Pl}^2$ we finally obtain the relation between $\MM$ and the measured Newton constant in the presence of a fifth force  \cite{Damour:1992we,polarski},
\begin{equation} \label{M-Mp}
G_{\rm N} = \frac{1}{8 \pi \MM^2 f} \, \frac{c + \MM^2 \dot f^2/f  }{c + \frac34 \MM^2 \dot f^2/f   }\, .
\end{equation}

\subsection{Mixing proportional to $\bar m^3_1$} \label{m31mix}

Here we consider in $S^{(2)}_{DE}$ of eq.~\eqref{example2} the operator $\bar m^3_1$ which can also produce kinetic mixing \cite{Creminelli:2008wc}. This operator is typically contained in galileon \cite{NRT,Deffayet:2009wt,Luty:2003vm,Nicolis:2004qq} and braiding models \cite{Deffayet:2010qz}, as we discussed in Secs.~\ref{sec_gal} and \ref{sec_braiding}.
We can read off the effect of the Stueckelberg trick on this operator from eqs.~\eqref{stuck1} and~\eqref{stuck4}. In particular, in Newtonian gauge at linear order we have
\be
\delta g^{00} \to -2 (\dot \pi - \Phi) \;, \qquad \delta K \to - (3 \dot \Psi + a^{-2} \nabla^2 \pi) \; .
\ee
Thus, by keeping only the kinetic terms the action reads 
\be
\begin{split}
S &= \int \sqrt{-g} \left(\frac{\MM^2}{2} R - c g^{00} - \frac{\bar m^3_1}{2} \delta g^{00} \delta K\right)\label{actionm13} \\
  & \!\!\!\!\!\overset{\rm kinetic}{\to}   \int  \MM^2 \left[ - 3 \dot \Psi^2 - 2 \vec \nabla \F \vec \nabla \PP + (\vec \nabla \PP)^2 \right] + c \dot \pi^2 - \tilde c (\vec \nabla \pi)^2 
  - 3 \bar m^3_1  \dot \Psi \dot \pi -  \bar m^3_1  \vec \nabla \Phi \vec  \nabla \pi    \, .
\end{split}
\ee
Here, as before, we have dropped the scale factor since we are considering scales much smaller than Hubble.
Also, we have defined 
\be
\tilde c \equiv c + \frac12 (H \bar m^3_1 + \dot{\bar m}^3_1 ) \;, \label{tildec}
\ee
where the second term on the right-hand side of this equation comes from integrating by parts the piece $ - a \bar m^3_1 \dot \pi  \nabla^2 \pi$.

It is handy to re-define the fields in such a way that they all have  dimensions of mass  as follows,
\begin{equation}
\pi_c \equiv c^{1/2} \pi\, , \quad \Psi_c \equiv \MM \Psi\, , \quad \quad \Phi_c \equiv \MM \Phi\, , \quad \alpha \equiv \frac{\bar m^3_1}{2 \MM c^{1/2}}\, \, .
\end{equation}
The Lagrangian becomes
\begin{equation}
{\cal L} = - 3 \dot \Psi_c^2 - 2\vec  \nabla \F_c \vec \nabla \PP_c + (\vec \nabla \PP_c)^2 + \dot \pi_c^2 - (\tilde c / c) (\vec \nabla \pi_c)^2  - 6 \alpha  \dot \Psi_c \dot \pi_c - 2 \alpha \vec   \nabla \Phi_c \vec \nabla \pi_c\, .
\end{equation}
In Fourier space, on the $(\Psi_c \Phi_c \pi_c)$ basis, the Lagrangian is a $3\times 3$ matrix,

\begin{equation}
{\cal L} \ \sim \ 
\begin{pmatrix} 
\Psi_c & \Phi_c  & \pi_c
\end{pmatrix}
\begin{pmatrix}
- 3 \omega^2 + k^2 \quad & - k^2  \quad & - 3 \alpha \omega^2 \\[2mm]
- k^2  &0& - \alpha k^2\\[2mm]
- 3 \alpha \omega^2 & -\alpha k^2 & \omega^2 - (\tilde c / c) k^2
\end{pmatrix} 
\begin{pmatrix} 
\Psi_c \\[2mm] \Phi_c \\[2mm] \pi_c
\end{pmatrix} \;. 
\end{equation}
The determinant evaluates
\begin{equation}
{\rm det} \, {\cal L} \ = \ - k^4 \left[    \omega^2 (1 + 3 \alpha^2)  - k^2 (\tilde c/c-\alpha^2)\right] \, ,
\end{equation} 
which corresponds to having one propagating degree of freedom with dispersion relation $\omega^2 \ = \ c_s^2 k^2$, where the sound speed $c^2_s$ is
\begin{equation}
c_s^2 \, =\, \frac{  \tilde c - \frac14 {\bar m^6_1}/\MM^2}{ c+\frac34  {\bar m^6_1}/\MM^2 }\, .
\end{equation}
Note that $c_s^2$ can become negative, in which case one has a gradient instability. We will get back to this point when we  study the stability in Sec.~\ref{sec_stability}.

One may wonder whether there exists a transformation of the metric that de-mixes $\pi$ from gravity, as in the example of the previous subsection. Indeed, it is easy to show that the two new potentials
\be \label{newpot}
\Phi_E = \Phi + \frac{\bar m^3_1}{2 \MM^2} \pi\,, \qquad  \Psi_E = \Psi +  \frac{\bar m^3_1}{2 \MM^2} \pi \; ,
\ee
kinetically de-mix the system. Comparing with eq.~\eqref{conf_met}, one sees that this cannot come from a conformal transformation of the metric.

As before, we can now add matter and study the Newtonian limit. We can neglect time derivatives and,  by varying the action \eqref{actionm13} with respect to $\Psi$, obtain $\Phi = \Psi$. That the two potentials  are equal  at linear order in the absence of anisotropic stresses is also evident from the traceless $ij$ part of the linearized  Einstein equations,
eq.~\eqref{eq_traceless} of App.~\ref{app_linear}. This is an example of a modification of gravity where, even in the absence of screening and in the presence of an explicit kinetic mixing, the post-Newtonian $\gamma$ parameter is unity. In this case, the action \eqref{actionm13} becomes
\begin{equation}
S = \int    - \MM^2 (\vec \nabla \F)^2 - \tilde c (\vec \nabla \pi)^2-  \bar m_1^3 \, \vec \nabla \F \vec \nabla \pi   - \F \delta \rho_m\, .
\end{equation}
Analogously to what we did in Sec.~\ref{sec_fdot} we derive the observed Newton constant, 
\begin{equation} \label{M-Mp2}
G_{\rm N} = \frac{1}{8 \pi \MM^2 } \, \frac{c + \frac12\left( \dot{\bar m}^3_1 + H \bar m_1^3\right)}{c  - \frac14 {\bar m_1^6}/{\MM^2} + \frac12\left( \dot{\bar m}^3_1 + H \bar m_1^3\right)  }\, ,
\end{equation}
where we have used the definition of $\tilde c$, eq.~\eqref{tildec}.
We stress again that the above analysis holds in the presence of a long-range propagating scalar degree of freedom and we have not assumed any screening mechanism.

\subsection{De-mixing in a quite general case and  stability}
\label{sec_stability}

Now we turn to a more general situation in which several of the discussed operators are switched on:
\begin{equation} 
S =  \int \sqrt{-g} \left(\frac{\MM^2}{2} f R - \Lambda - c g^{00} + \frac{M_2^4}{2} (\delta g^{00})^2  - \frac{\bar m^3_1}{2} \delta g^{00} \delta K  \right) + \frac12 \sqrt{-g}T^{\mu \nu} \delta g_{\mu \nu}\;. \label{acsecondein}
\ee
Here,   in the last term, we have also added  the coupling of matter to gravity. This will give us the opportunity to discuss the effect of fifth-force mediation on the Newton constant.  Note that for the present discussion $\Lambda$ is irrelevant.

Once a DE model is translated in our framework and ``distilled" into a certain number of operators, it is straightforward to address its stability by de-mixing $\pi$ from gravity and looking at its quadratic action.  Only $f$ and $\bar m_1^3$ produce kinetic mixing in the above action. We should then combine the linear transformations~\eqref{conf_met} and~\eqref{newpot} and define 
\be 
\label{newpot2}
\begin{split}
\Phi_E &= \Phi + \frac12\bigg(\frac{\dot f}{f} + \frac{\bar m^3_1}{\MM^2} \bigg) \pi\,, \\
\Psi_E &= \Psi -  \frac12\bigg(\frac{\dot f}{f} - \frac{\bar m^3_1}{\MM^2} \bigg) \pi \;\, .
\end{split}
\ee
The kinetic quadratic piece of action~\eqref{acsecondein} written on this new basis features a normal non-propagating gravity sector (see eq.~\eqref{normal_gravity}) plus the following de-mixed scalar piece:
\begin{multline} \label{pai}
S_\pi \overset{\rm kinetic}{=} \int \, a^3 \Bigg\{ \bigg[c + 2 M_2^4 + \frac34 \frac{\dot f^2}{f} \MM^2 - \frac32 \bar m_1^3 \frac{\dot f}{f} + \frac34 \frac{\bar m_1^6}{\MM^2 f} \bigg] \dot \pi^2 \\ - \bigg[c + \frac34 \frac{\dot f^2}{f} \MM^2 - \frac12 \bar m_1^3 \frac{\dot f}{f} - \frac14 \frac{\bar m_1^6}{\MM^2 f}  + \frac12\left( \dot{\bar m}^3_1 + H \bar m_1^3\right) \bigg] \frac{(\vec \nabla \pi)^2}{a^2} \Bigg\}\, . 
\end{multline}
The condition that there are no ghost instabilities requires that the brackets in the first line of this equation is positive. Note that by eq.~\eqref{c2} $c = (\rho_{D}+p_{D})/2 -  (\ddot f  -  H \dot f ) \MM^2   $. Thus, by an appropriate choice of parameters the equation of state may become phantom, $p_D<-\rho_D$, and yet be ghost free---for instance, when $M_2^4$ dominates the kinetic term \cite{EFT1,Creminelli:2008wc}, in the presence of a time dependent $f$ \cite{Das:2005yj} or of a non-vanishing $\bar m_1^3$ \cite{Deffayet:2010qz}.

The square of the speed of sound of fluctuations is given, as usual, by the ratio of the brackets in the second line of action \eqref{pai} to the one in the first line, i.e.,
\be \label{cies}
c_s^2 = \frac{c + \frac34  \MM^2 {\dot f^2}/{f} - \frac12 \bar m_1^3 {\dot f}/{f} - \frac14 {\bar m_1^6}/( {\MM^2} f)  + \frac12\left( \dot{\bar m}^3_1 + H \bar m_1^3\right) }{c + 2 M_2^4 + \frac34  \MM^2 {\dot f^2}/{f} - \frac32 \bar m_1^3 {\dot f}/{f} + \frac34 {\bar m_1^6}/ ( {\MM^2} f) }\;.
\ee
For $\bar m_1^3=0$ we reproduce the speed of sound computed in \cite{Sawicki:2012re} for a non-minimally coupled scalar field. 
Since the denominator on the right-hand side is positive by the ghost-free condition, $c_s^2$ has the same sign as the numerator. This may  become negative, in which case one has a gradient instability. In particular, regardless of the sign of $\bar m_1^3$ the mixing with gravity always induces  
a Jeans-like instability  proportional to $\bar m_1^6$. This instability is discussed in \cite{EFT1,Creminelli:2008wc}, where it is assumed that $M_2^4 $ dominates the time kinetic term  and that $f$ and $\bar m_1^3$ are constant, in which case it is cured by requiring that $ \bar m_1^3/\MM^2 \lesssim H$. More generally, ensuring that the time for these instabilities to develop must be longer than the age of the Universe constrains  the parameters of  action \eqref{ac}.

The calculation of the modified Newton constant for this DE theory can be done straightforwardly using the methods developed in the previous two subsections. It yields
\begin{equation} \label{M-Mp_total}
G_{\rm N} = \frac{1}{8 \pi \MM^2 f} \, \frac{c +  \MM^2 \dot f^2/f + \frac12\left( \dot{\bar m}^3_1 + H \bar m_1^3\right)}{c + \frac34 \MM^2 \dot f^2/f   - \frac12 \bar m_1^3 {\dot f}/{f} - \frac14 {\bar m_1^6}/ ({\MM^2} f) + \frac12\left( \dot{\bar m}^3_1 + H \bar m_1^3\right)  }\, .
\end{equation}

\section{Matching with DE models}
\label{sec:matching}

To show that the action \eqref{ac} is very general and encompasses many of the models of the literature, here we discuss how one can translate in the EFT language several well-known examples. 

\subsection{Quintessence and $k$-essence} \label{sec_quint}

For $f(t)=1$, the action \eqref{ac} contains any minimally-coupled single-field dark energy model \cite{Creminelli:2008wc}. For instance, as discussed in the introduction, a model with minimal kinetic term such as standard quintessence  is contained in its first  line. Indeed, its Lagrangian can be rewritten in unitary gauge as 
\be
- \frac12 (\partial \phi)^2 - V(\phi) \to  - \frac12 \dot \phi_0^2(t) g^{00}- V(t)\;,
\ee
which is reproduced by the first line of \eqref{ac}, once we note that $2 c(t) = \rho_D  + p_D  = \dot \phi_0^2(t)$ and $2 \Lambda(t) = \rho_D -p_D =  2 V(t)$.

Similarly, the action \eqref{ac} includes also $k$-essence models \cite{ArmendarizPicon:2000dh}, where the Lagrangian has at most one derivative acting on the field, ${\cal L} = P(\phi, X)$. In unitary gauge this is of the form $P( \phi_0(t) , \dot \phi_0^2 g^{00})$, which can be expanded in powers of $\dot \phi_0^2  \delta g^{00}$ and written in the form \eqref{ac} \cite{EFT2,Creminelli:2008wc}. Note that one can always make a field redefinition such that $\phi_0 = t$. In this case the action  \eqref{ac} is reproduced with
\be
c(t) - \Lambda(t)  = P  (t) \;, \quad
c(t)  =  \left. \frac{\partial P}{\partial X} \right|_{X=1}\; ,  \quad
M_n^4 (t)  =  \left. \frac{\partial^n P}{\partial X^n} \right|_{X=1}  \quad (n\ge 2) \;.
\ee

\subsection{DGP and galileon} \label{sec_gal}

Terms containing the extrinsic curvature can  become important around localized gravitational sources and be responsible for screening mechanisms, such as the Vainshtein effect \cite{Vainshtein:1972sx}, which allow models based on light fields with gravitational-strength coupling to evade Solar System constraints.
In particular, in the decoupling limit the DGP model \cite{Dvali:2000hr} is described by a non-minimally coupled scalar field with cubic self-interaction \cite{Luty:2003vm,Nicolis:2004qq},
\be
(\partial \phi)^2 \Box \phi \;, \label{DGP}
\ee
which is the prototypical operator responsible for the Vainshtein mechanism and which is also the cubic operator ${\cal L}_3$ of the galileon \cite{NRT}.
Now we want to rewrite this operator in unitary gauge to show that it  is actually contained in  action \eqref{ac}.
Specifically, we can expand the operator \eqref{DGP}  in powers of $\delta g^{00}$ and at most one power of $\delta K$ (see also \cite{Creminelli:2010ba}). 

Using the definition of the extrinsic curvature, eq.~\eqref{definition_ext}, we note that, up to a boundary term, the density Lagrangian can be rewritten as
\be
(\partial \phi)^2 \Box \phi =  \frac23 \left[- (\partial \phi)^2\right]^{3/2} K\;.
\ee
In unitary gauge this becomes
\be
\begin{split}
(\partial \phi)^2 \Box \phi   \to  & \ (2/3) (-g^{00})^{3/2} \dot \phi_0^3  K\\
& = - 2 \dot \phi_0^2  \left[  \ddot \phi_0 \sqrt{- g^{00}} -  H \dot \phi_0 \left( (-g^{00})^{3/2} -1 \right) \right]  + (2/3) \left[ (-g^{00})^{3/2} -1 \right]\dot \phi_0^3  \delta K \;,
\label{DGP_exp}
\end{split}
\ee
where for the second line we have used $K = 3 H + \delta K$ and we have integrated by parts $(2/3) \dot \phi^3 K$ using
\be
K= - \nabla_\mu \left( \frac{g^{\mu 0}}{\sqrt{-g^{00}}}  \right)\;.
\ee
Thus, the DGP operator \eqref{DGP} can be written in the EFT language  as a polynomial of $\delta g^{00}$ and $\delta K$ with at most one power of $\delta K$.

As a concrete example of a model which contains the operator \eqref{DGP}, let us consider the galilean cosmology discussed by Chow and Khoury in Ref.~\cite{justin} and rewrite it in our EFT language. The full action in Jordan frame reads (see eq.~(3.1) of this reference)
\be
S =  \int d^4 x \sqrt{-g} \left[  \frac{\MM^2}{2} e^{-2 \phi/\MM} R - \frac{r_c^2}{\MM}(\partial \phi)^2 \Box \phi + {\cal L}_m \right]\;, \label{action_galileon}
\ee
where $r_c$ is a length-dimension parameter.
We can expand this action in series of $\delta g^{00}$ and $\delta K$ by making use of eq.~\eqref{DGP_exp}. Comparing with \eqref{ac} we find,
\begin{align} 
f(t) & = e^{-2 \frac{\phi_0}{\MM}}\ \;, \quad
\Lambda (t)  = -\frac{r_c^2}{\MM} \, \dot \phi_0^2( \ddot \phi_0 + 3 H \dot \phi_0)\;, \quad 
c (t)=  \frac{r_c^2}{\MM}\, \dot \phi_0^2(\ddot \phi_0 - 3 H \dot \phi_0) \;, \label{cc}\\
M_2^4 (t)& =  -\frac{r_c^2}{2 \MM}\, \dot \phi_0^2(\ddot \phi_0 + 3 H \dot \phi_0)\;,\quad M^4_3 (t)= -  \frac{3r_c^2}{ 4\MM}   \dot \phi_0^2(\ddot \phi_0 + H \dot \phi_0) \;, \quad \text{etc.} \;,\\
\bar m_1^3 (t)& = -\frac{r_c^2}{\MM} 2 \dot \phi_0^3\;, \quad \bar m_2^3 (t) =  \frac{r_c^2}{2 \MM}  \dot \phi_0^3 \; , \quad
\text{etc.} \;, \label{endend}
\end{align}
where we have stop the comparison at third order in the perturbations. Note that terms containing more than one power in the perturbation of the extrinsic curvature do not appear, so that
$\bar M_2 (t)=  \bar M_3 (t)= \ldots =0$.

As an application of the machinery of Sec.~\eqref{sec_stability}, it is possible to study the stability of this model. Our analysis is slightly different from that of \cite{justin}, as it takes into account the kinetic mixing with gravity. Using eq.~\eqref{pai}, the requirement that there is no ghost instability turns into 
\begin{equation}
-2\frac{r_{c}^{2}}{\MM}\dot\phi_{0}H+\frac{r_{c}^{4}}{\MM^{4}}\dot\phi_{0}^{4}-2\frac{r_{c}^{2}}{\MM^{2}}\dot\phi_{0}^{2}+e^{-2\frac{\phi_{0}}{\MM}}>0 \;,
\end{equation}
while the speed of sound squared,  replacing eqs.~\eqref{cc}--\eqref{endend} into  eq.~\eqref{cies}, reads
\begin{equation}
c_{s}^{2}=\frac{- \frac43 H \dot\phi_{0}r_{c}^{2}/\MM-\frac23 \dot\phi_{0}^{2}r_{c}^{2}/\MM^{2}-\frac13 \dot\phi_{0}^{4}r_{c}^{4} e^{2\phi_{0}/\MM} /\MM^{4}- \frac23 \ddot\phi_{0}r_{c}^{2}/\MM+e^{-2\phi_{0}/\MM}}{-2H\dot\phi_{0}r_{c}^{2}/\MM-2 \dot\phi_{0}^{2}r_{c}^{2}/\MM^{2}+\dot\phi_{0}^{4}r_{c}^{4} e^{2\phi_{0}/\MM} /\MM^{4}+ e^{-2\phi_{0}/\MM}} \;.
\label{ss_justin}
\end{equation}
Using the approximate solution studied in \cite{justin}, $\dot \phi_0 \approx \pm (2/3)^{1/2} \MM r_c^{-1}$, $H\approx (2/3)^{3/2} r_c^{-1}$ and $\phi_0 \ll \MM$, we confirm that the stable branch is for $\dot \phi_0 <0$, for which the sound speed squared \eqref{ss_justin} exactly reduces to $c_s^2 =1$, a different result from that of \cite{justin}, i.e.~$c_s^2=2/3$.

\subsection{Kinetic braiding} \label{sec_braiding}

It is straightforward to show that the operator 
\begin{equation}
G(\phi, X) \square \phi \;,
\end{equation}
denoted in \cite{Deffayet:2010qz} as kinetic braiding, can also be rewritten as a polynomial of $\delta g^{00}$ and $\delta K$ with at most one power of $\delta K$.
We can expand $G(\phi, X)$ in polynomials of $\delta X$, say,
\begin{equation}
G(\phi, X)= \sqrt{-X} \, \sum_{m=0}^\infty l_m (\phi)  \delta X^m,
\end{equation}
where the $\sqrt{-X}$ on the right-hand side helps simplifying the notation and $l_m(\phi)$ are the coefficients of the expansion. Then we can use the relation
\be
\square \phi = - \sqrt{-X} K + \frac{1}{2 \sqrt{-X}} n^\mu \partial_\mu X\; .
\ee
In unitary gauge $X = \dot \phi_0^2 (-1 + \delta g^{00})$ and we will take $\dot \phi_0 = 1$ for simplicity. Terms linear in $K$ can be integrated by parts as in \eqref{80A} and give powers of $\delta g^{00}$. After few integrations by parts we finally obtain
\be
G(\phi, X)  \square \phi \ =\ \dot l_0(t) \sqrt{-g^{00}} - \sum_{m=1}^\infty \left( l_m (t) K + \frac{1-2 m}{2m } l_{m-1} (t)K + \frac{\dot l_{m-1}(t)}{2 m} \sqrt{-g^{00}}\right) (\delta g^{00})^m  \;.
\ee
The contributions of such an operator to action~\eqref{ac} read
\begin{align} 
\Lambda (t)  &= 3 H \left(l_1 - \frac{1}{2} l_0\right) \;, \quad 
c (t)=  \dot l_0 + 3 H \left(l_1 - \frac{1}{2} l_0\right)  \\
M_2^4 (t)& = \frac{\dot l_0}{4} - \frac{\dot l_1}{2} - 6 H l_2 + \frac{9}{2} H l_1,\quad M^4_3 (t)= \ldots \;,\\
\bar m_1^3 (t)& = 2 l_1 -  l_0, \quad \bar m_2^3 (t) = \ldots \;. 
\end{align}

\subsection{Ghost condensate and khronon} \label{sec_khro}

Operators containing $\delta K^2$, $(\delta K_{\ \mu}^\nu)^2$ and gradients of $g^{00}$ have more than one derivative acting on a scalar. These terms can become relevant in models where the spatial gradients are parametrically more important than the time derivatives. 

For instance, the ghost condensate theory \cite{ghost}, based on the field shift symmetry $\phi \to \phi +\,$const, is realized in the limit of $(\rho_D +p_D ) \to 0$, i.e.~$c(t)\to 0$---here we are assuming minimal coupling, $f=1$. In this limit the spatial kinetic term of the perturbations is dominated by the higher-derivative operators proportional to $\bar M_2$ and $\bar M_3$ \cite{ghost,EFT1,EFT2,Creminelli:2008wc}. 
More generally, the limit of zero sound speed for models with $p_D \neq -\rho_D$ can be obtained as a small deformation of the ghost condensate theory  \cite{Creminelli:2009mu,Sefusatti:2011cm,D'Amico:2011pf} (see however \cite{Lim:2010yk} for another way of reproducing this limit). In this context, absence of ghost and gradient instabilities can be guaranteed also for an  equation of state $p_D  < - \rho_D $ by the presence of these higher-derivative operators \cite{EFT1}, even though their effect is practically absent on cosmologically relevant scales \cite{Creminelli:2008wc}.

Another example is the khronon field, based on the full reparametrization invariance $\phi \to \tilde \phi(\phi)$. This symmetry has recently received attention in the context of Horava gravity and its healthy extensions \cite{Horava:2009uw,Blas:2009qj,Blas:2010hb} and it has been used to construct a technically natural dark energy model in \cite{Blas:2011en} and test Lorentz invariance of dark matter \cite{Blas:2012vn}. The invariant object under this symmetry is the unit vector perpendicular to the constant time hypersurfaces $n^\mu$---defined in \eqref{nmu}---and to lowest order in derivatives the action can  be written as \cite{Blas:2010hb,Creminelli:2012xb}
\be
S =\frac12 \int d^4 x \sqrt{-g} \left( \MM^2 R - 2 \Lambda - M_\lambda^2 (\nabla_\mu n^\mu - 3H)^2 + M_\alpha^2 (u^\mu \nabla_\mu u^\rho)^2 \right)\;,
\ee
where $M_\lambda$ and $M_\alpha$ are the two mass parameters of the model, beside the vacuum energy $\Lambda$. It is straightforward to show that in unitary gauge this action is equivalent to \eqref{ac} with\footnote{To rewrite the last term, use eq.~(73) of Ref. \cite{EFT2}.}
\be
\begin{split}
\Lambda (t) & = \Lambda\; , \quad
c(t)=0\;,\\
\bar M_2 (t)& = M_\lambda\;, \quad
m_2 (t) = \frac{M_\alpha}{2} \;.\\ 
\end{split}
\ee
Thus, symmetry under reparametrization $\phi \to \tilde \phi (\phi)$ imposes that all the terms without derivatives on $g^{00}$ are absent, i.e.~$c(t)=M_i (t) =0$.

\subsection{$F(R)$ gravity} 
\label{sec_f(r)}

Curvature invariants offer an interesting application of our formalism. We should distinguish two types of situations. At the low energy effective level, the gravitational Lagrangian does include a series of terms made of curvature invariants, such as $R^2$, $R_{\mu\nu}R^{\mu\nu}$ etc., weighted by the smallest masses that have been  integrated out (see e.g.~the nice review~\cite{cliff}). These terms do not generate new degrees of freedom  and their natural size is generally way too small to be relevant for the DE problem.  Our formalism can encode curvature invariants; as an example, in App.~\ref{sec_invariants} we consider the covariant operator $A(\phi) R_{\mu \nu}R^{\mu \nu}$ and calculate its contributions to the various terms of action~\eqref{ac}. 

When higher-order curvature invariants are used, naively,  beyond the EFT regime, they normally lead to ghost instabilities~\cite{woodard}. As far as we understand, the only exceptions are general function of the Ricci scalar $R$~\cite{Starobinsky:1980te,Capozziello:2003tk,Carroll:2003wy} and of the Gauss-Bonnet term~\cite{Nojiri:2005jg}. Leaving aside naturalness problems, there has been a lot of activity in trying to engineer DE models using Lagrangians of the $F(R)$ type (see e.g.~\cite{DeFelice:2010aj} for a review and further references). The equivalence between $F(R)$ and Brans-Dicke theories is well known (see for instance \cite{Chiba:2003ir}). Due to the strong coupling of the resulting scalar-tensor theory, the Chameleon mechanism~\cite{chame} is needed to make $F(R)$ models phenomenologically acceptable.

Let us see how to treat $F(R)$ theories within our formalism. First we should note that a generic expansion in $\delta R$ is allowed by the set of rules that we explained in Sec.~\ref{sec_secondorder}. Indeed, by expanding around a FRW solution with Ricci scalar $R^{(0)} (t)$, we obtain the second line of eq.~\eqref{expansionfR}.
This is already a legitimate DE action in unitary gauge: as discussed in Sec.~\eqref{sec_hidden}, at quadratic order it generates a propagating degree of freedom with relativistic dispersion relation. However, if we attempt the Stueckelberg trick at this point, we find that no $\pi$ is indeed produced. 
This is not surprising: neither $R$, nor any other curvature invariant produce $\pi$ field upon the infinitesimal time diffeomorphism $t\rightarrow  t+\pi(x)$, just because they are gauge invariant.

However, we can   fix the time slicing in such a way that it coincides with uniform $ R $ hypersurfaces. 
Note that this does not mean that we are killing the dynamics, because $\delta R$ has a nontrivial structure when written in terms of metric components. For instance, the linear term in $\delta R$ in the first line of eq.~\eqref{expansionfR} still produces non-trivial equations of motion when varied with respect to some metric component. What this prescription does kill are all the terms beyond the linear order in the expansion~\eqref{expansionfR}, because their contributions to the equations of motion have always at least one power of $\delta R$, and therefore vanish. 
Finally, by choosing the time such that $R^{(0)} = t$ from eq.~\eqref{expansionfR} we obtain the unitary-gauge action
\begin{equation}
\label{ac_fR}
\int d^4 x \sqrt{-g} \frac{\MM^2}{2} F(R) \ \rightarrow \ \int d^4 x \sqrt{-g} \frac{\MM^2}{2} \left[ F'(t) R + F(t) - t   F'(t) \right]\, .
\end{equation}
This action is equivalent to \eqref{ac} upon identifying
\be \label{chiba}
\begin{split}
f(t) & = F'(t) \;, \quad
\Lambda (t)  = - \frac{\MM^2}{2} \left(F(t ) -  t  F'(t) \right) \;,\quad
c(t)   =0\;,\\
M_i & = \bar M_i = \ldots =0\;.
\end{split}
\ee
Since now the gauge is fixed, $\pi$ terms are produced by expanding the time dependent functions $F(t)$ and $F'(t)$. At first sight, $\pi$ is non-dynamical because it does not have a kinetic term. However, the latter can be generated by de-mixing the field from gravity, as explained in Sec.~\ref{sec_fdot}. Indeed, as expected, action \eqref{ac_fR}  is the one of a scalar field without kinetic term in Jordan frame, written in unitary gauge and the above procedure is alternative and equivalent to the standard trick that introduces a scalar field $\phi$ in the action~\cite{Chiba:2003ir}.

\section{The ``Einstein Frame"} \label{sec_einst}

The constructive approach explained in the introduction naturally leads to a Jordan frame formulation of DE in the unitary gauge. The Jordan metric has also the advantage of being uniquely defined by the coupling to matter and more directly related to observables as we noted in Sec.~\ref{sec_general_1}. When $f$ is not constant, however,  the gravitational dynamics looks obscure in the Jordan frame, just because the new scalar degree of freedom, $\pi$, mixes with the gravitational ones---say, $\F$ and $\PP$ in Newtonian gauge. In the absence of kinetically mixing quadratic operators such as $\bar m_1^3$, the system can be diagonalized in a very standard way by a conformal redefinition of the metric tensor, i.e., by going to the so called Einstein frame, which is the subject of this section. However, one should keep in mind that, in the presence of mixing quadratic operators, gravity cannot be de-mixed in this way, although it is still possible to diagonalize the quadratic kinetic action as shown in Sec.~\ref{sec_mixing}.  

In this section  we adopt the standard notation of scalar-tensor theories and call ``Einstein frame" the conformally related metric in which the gravitational action does not contain a function multiplying the Einstein-Hilbert term and in which there is no mixing proportional to $\dot f$. 
Instead of just conformally transforming all our operators and coefficients, in the next subsection we provide an alternative, ``Einstein frame" construction of our formalism as a natural extension of the EFT of quintessence developed in~\cite{Creminelli:2008wc}.  We then re-switch back to Jordan frame in Sec.~\ref{sec_etoj}.

\subsection{EFT of DE in Einstein frame: alternative construction\footnote{This subsection and App.~\ref{app:field_red} are the only places in the paper where Einstein-frame quantities \emph{are not} denoted with a hat nor with the subscript $E$ and Jordan-frame quantities explicitly carry a subscript $J$.}} \label{sec_general}

It was shown in \cite{EFT2} that the most generic theory with broken time diffeomorphisms around a given FRW background can be written as 
\be
\label{EFTI}
S=   \int d^4x \sqrt{-g} \left[ \frac{ \MM^2 R}{2}  - c g^{00} -  \Lambda + F^{(2)} (\delta g^{00}, \delta K_{\mu \nu}, \delta R_{\mu \nu \rho \sigma}; h^{\ \mu}_{\nu} \nabla_\mu ; t ) \right]\;,
\ee
where $c$ and $\Lambda$ are  fixed by the background evolution, 
\be
c(t) = - \MM^2 \left( \dot H - \frac{k}{a^2} \right)\;, \quad \Lambda (t) = \MM^2 \left( 3 H^2 + \dot H + 2 \frac{k}{a^2} \right)\;,
\ee 
and $F^{(2)}$ is a quadratic or higher-order function of its arguments, which generates  terms  invariant under spatial diffeomorphisms and of quadratic or higher order in the fluctuations around a given FRW background \cite{EFT1}. 

Action \eqref{EFTI} was used to describe the EFT of inflation in the context of single field models. One of the main differences of the current phase of cosmic acceleration with inflation is the presence of matter species such as cold dark matter, baryons, radiation and neutrinos. In this case, one must supplement eq.~\eqref{EFTI} with an action for these components. Neglecting mutual interactions between components, which are irrelevant for late-time cosmology, the action reads
\be
\label{EFTQ}
S=   \int d^4x \sqrt{-g} \left[ \frac{R \MM^2}{2}  - c g^{00} -  \Lambda + \ldots \right] + \sum_i S_{m,i} \left[ g_{\mu \nu}, \psi_{i} \right] \;,
\ee
where the index $i$ runs over all matter species and the ellipses stand for higher-order terms.
The functions $c$ and $\Lambda$ can be fixed imposing 
that the background  energy density and pressure of the dark component $\rho_D $ and $p_D $ satisfy the Friedmann equations. More specifically, the background Einstein equations derived from this action read
\be
\label{Ein_Ein1}
G_{\mu \nu} \MM^2 + (   \Lambda - c)  g_{\mu \nu} -2 c  \delta^0_\mu \delta^0_\nu = \sum_i T^{(i)}_{\mu \nu}\;.
\ee
Thus, using~\eqref{backvalues} in this equation yields the Friedmann equations with  
\be
\label{cL}
c(t)= \frac12 \left( \rho_D  + p_D  \right) \;, \quad \Lambda(t)= \frac12 \left( \rho_D  - p_D  \right) \;.
\ee
Once the linear terms have been fixed using \eqref{cL},  action \eqref{EFTQ} describes the most generic  theory of single-field dark energy perturbations around a give FRW background evolution. We stress that in this action the higher-order terms in the dots are constructed from the metric and from geometrical objects describing the preferred time-slicing  and do not explicitly involve matter quantities, such as the matter energy-momentum tensor. Thus, matter is minimally coupled and for this reason this theory has been dubbed  EFT of quintessence \cite{Creminelli:2008wc}.

To go beyond the assumption of minimal coupling, we can couple  each matter species  to a metric $g^{J (i)}_{\mu \nu}$ which differs from the Einstein frame metric, now denoted by $ g_{\mu \nu}$, for which the mixing between dark energy and gravity is minimal. Schematically,  
\be
S_{m,i} [g_ {\mu \nu}, \psi_i] \ \longrightarrow S_{m,i} \big[  g^{J(i)}_{\mu \nu}, \psi_{i} \big] \;.
\ee

For each matter species, the metric $g^{J(i)}_{\mu \nu}$ is a two-index tensor whose most generic form can be given as a function of the Einstein frame metric $g_{\mu \nu}$ and of geometrical objects invariant under spatial diffeomorphisms, constructed in the Einstein frame. We will assume for simplicity  that all species couple to the DE sector in the same way, but our arguments can be straightforwardly extended  to considering non-universal couplings\footnote{The weak equivalence principle is so well constrained (the level is $\lesssim 10^{-13}$~\cite{Schlamminger:2007ht}) that its possible tiny violations are hardly relevant for DE---this is why it is a good approximation to assume a universal coupling here. On the other hand, couplings to the \emph{dark matter} sector are far less constrained (order $10 \%$~\cite{constr1,constr2}) and play a crucial role in models of coupled quintessence (e.g.~\cite{Amendola:1999er,Gasperini:2001pc,Comelli:2003cv}). It would be interesting to extend our formalism to those scenarios.}.  In unitary gauge the Jordan metric takes generically the form
\begin{align}
g_J^{\mu \nu} = & \ f( t)  g^{\mu \nu}  +   g^{\mu \nu} \left[ \beta_1 ( t) \delta  g^{00} + \frac12 \beta_2 ( t) (\delta  g^{00})^2 + \dots  \right] +  \delta  g^{0 \mu} \delta  g^{0\nu} \left[\tilde\gamma_0( t) + \tilde\gamma_1 ( t) \delta  g^{00}  + \dots  \right] \nonumber \\
&+ \frac12 \tilde m_1^{-1}( t)  g^{\mu \nu} \delta g^{00} \delta  K
+ \frac12 \tilde m_{2}^{-2}( t) \delta  K^{\mu \rho} \delta  K_{\rho}^{\ \nu} + \ldots\;,\label{JE}
\end{align}
where $f( t)$, $\beta_i( t)$, $\tilde \gamma_i ( t)$ and $\tilde m_i( t)$ are functions of time. The first line of this equation includes conformal and disformal transformations \cite{Bekenstein:1992pj}, allowing for kinetic mixing. Note that the Jordan metric has been written in such a way that the only  term contributing to its background is the first on the right-hand side, $f( t)  g^{\mu \nu}$;  the only \emph{linear} term is the one proportional to $\beta_1( t)$, while the other terms are higher order in the perturbations. Their exact form is irrelevant for our discussion. Indeed, in App.~\ref{app:field_red} we show that all the terms on the right-hand side of~\eqref{JE} except the first can be reabsorbed by a perturbative field redefinition, so that, for what concerns the matter action, we can write\footnote{A similar field redefinition to rewrite the action in a minimal form has been also done in \cite{BF}, see e.g.~their equation (1.2).}
\be
S_m[g^{\mu \nu}_J,\psi_m] = S_m[f( t)  g^{\mu \nu},\psi_m]\, .
\ee
This field redefinition leaves  the function $  \Lambda $ unchanged, while changing $ c $ and  the time-dependent coefficients of the higher-order terms. However, the overall structure of the gravitational/DE sector action remains the same as eq.~\eqref{EFTI}.
In summary,  
 \be
\begin{split}
\label{EFTDE_E2}
S=   & \int d^4  x \sqrt{- g} \left[ \frac{ R \MM^2}{2}  -  c g^{00} -   \Lambda \right. \\
&+  \left.\frac{M_2^4}2 (\delta g^{00})^2 + \frac{M_3^4}{3!} (\delta g^{00})^3 + \ldots 
  - \frac{\bar m_1^3}{2}\,  \delta g^{00} \delta K - \frac{\bar M_2^2}{2}  \delta K^2  + \dots
  \right]
\\ & +  S_{m} \left[ f^{-1} ( t)  g_{\mu \nu}, \psi_{i} \right] \;,  
\end{split}
\ee
where in the second line we have explicitly written some of the higher-order terms.

Before deriving the Jordan frame action, let us briefly review the background evolution equations in the Einstein frame. In this frame, the  background Einstein's equation is not affected by the presence of the conformal function $f(t)$ inside the matter action and it is still given by 
\be
\label{Ein_Ein2}
G_{\mu \nu} \MM^2 + (   \Lambda - c ) g_{\mu \nu} -2 c  \delta^0_\mu \delta^0_\nu =  T_{\mu \nu}\;.
\ee
 Thus, eq.~\eqref{cL} still holds. However, the background matter energy-momentum tensor in the  Einstein frame is related to the one in the Jordan frame by
$T^E_{\mu \nu} = f^{-2} T^J_{\mu \nu}$, which implies $\rho^E_m \equiv \rho^J_m f^{-2}$ and $p^E_m \equiv p^J_m f^{-2}$. 
The Friedmann equations are obtained from~\eqref{Ein_Ein2} and are the usual ones,
\begin{align}
  H^2 + \frac{k}{ a^2}    &=  \frac1{3  \MM^2} ( \rho_m +  \rho_{D}  )  \; ,\\
\dot { H} -  \frac{k}{  a^2} &=  - \frac1{2  \MM^2} ( \rho_m +  \rho_{D} + p_m +  p_{D}  )  \;.
\end{align}
As expected, matter and dark sector are coupled in this frame and their conservation equations read
\begin{equation}
\dot{ \rho}_m + 3  H ( \rho_m +  p_m) = \frac12 \frac{\dot f}{f}  T_m \;, \qquad
\dot{ \rho}_D + 3  H ( \rho_D  +  p_D ) = - \frac12 \frac{\dot f}{f}  T_m \;.
\end{equation}

\subsection{Going to Jordan frame} \label{sec_etoj}

To go to the Jordan frame we perform a conformal transformation $d  s_E^2 = f d  s_J ^2$ in the action \eqref{EFTDE_E2}. In this subsection, contrarily to what we did in the previous one, we drop the index $J$ from Jordan frame (or ``J-frame", for brevity) quantities and we use the index $E$ for Einstein frame (or ``E-frame") quantities. At times, E-frame quantities are also denoted by a hat to make the notation more compact. We have
\begin{align}
\label{ct_1} d^4  x_E  \sqrt{- g_E} & = d^4 x f^2 \sqrt{-g}  \;, \\
R_E &= f^{-1} \left(  R - 3  \square \ln f - \frac32  g^{\mu \nu}  \partial_\mu \ln f  \partial_\nu \ln f \right) \;.
\end{align}
Things simplify by rescaling the time and the scale factor such that $d t_E = f^{1/2} dt$ and $a_E = f^{1/2} a$, while leaving spatial coordinates unchanged. With this choice,  in unitary gauge the  conformal transformation of  the metric components reads\footnote{Note that  these relations only hold in unitary gauge. Some care has to be taken when extending them to an arbitrary gauge. For instance, the conformal transformation $d  s_E^2 = f d  s_J ^2$ in unitary gauge reads $g^E_{00} d t_E^2 = f g_{00} dt^2$, which in a general gauge becomes, at linear order,
$g^E_{00} d t_E^2 = f \big( 1 + (\dot f/f) \pi \big) g_{00} dt^2$.  
In Newtonian gauge this yields $\Phi_E = \Phi + (\dot f/f) \pi $, which is of course consistent with eq.~\eqref{conf_met}.}
\be
\label{metric_transf}
g^E_{00} =   g_{00}\;, \qquad
g^E_{0i} =   (a_E /a) g_{0i}\;, \qquad
g^E_{ij} =   (a_E /a)^2 g_{ij}\;,
\ee
and simplifies the transformations of the unitary-gauge quantities in the action.

It is convenient to first discuss how the first and last lines of the action \eqref{EFTDE_E2} transform, discarding for the moment the higher-order terms in the second line and leaving the discussion of their conformal transformation to the end of this section. Indeed, as expected, transforming the higher-order terms does not affect the first and last lines of the action. In this case, the action in Jordan frame reads as in eq.~\eqref{example2}, 
with $c$ and $\Lambda$ related to the analogous E-frame quantities by
\be
\label{c_L_EJ} c= f^{2} \left[ c_E - \frac34 \left( \frac{d \ln f}{d t_E} \right)^2 \!\MM^2 \right] \;, \qquad
\Lambda = f^{2} \Lambda_E\;.
\ee
The J-frame background energy density and pressure defined  in~\eqref{frie1}--\eqref{frie2} are related to their E-frame counterparts introduced in the last subsection by
\begin{align}
\rho_D  & = f^2  \left[ \rho^E_D  + 3 \MM^2 \frac{d \ln f}{d t_E} \left(  \frac14 \frac{d \ln f}{dt_E} - H_E \right) \right]\;, \\
p_D  & = f^{2} \left[ p^E_D   + \MM^2 \left(\frac{d^2 \ln f}{dt_E^2} -\frac14 \left(\frac{d \ln f}{dt_E}\right)^2 +2 \frac{d \ln f}{d t_E}  H_E  \right) \right]\;.
\end{align}
Note that the Hubble parameter is different in the two frames,
\begin{equation}\label{hubbles}
H_E = f^{-1/2}\bigg(H + \frac12 \frac{\dot f }{f}\bigg)\; ,
\end{equation}
where a dot denotes a derivative with respect to the J-frame time.

For the higher-order terms in the action, let us start by discussing the polynomials of $\delta g^{00}$. 
Since $\delta g^{00}$ is not affected by a conformal transformation in unitary gauge (see eq.~\eqref{metric_transf}), the coefficients of this expansion will be simply multiplied by the factor $f^2$ coming from the transformation of the volume element, eq.~\eqref{ct_1}. Thus, in Jordan frame the first two terms of the second line of eq.~\eqref{EFTDE_E2}
simply become
\be
d^4 x\sqrt{- g} \left[  \frac{ f^2 M_2^4}{2!} (\delta g^{00})^2 +  \frac{f^2 M_3^4}{3!}  (\delta  g^{00})^3 + \ldots  \right]\;. \label{ac2}
\ee
We can now discuss the terms containing the extrinsic curvature. From the definition of the extrinsic curvature and using eq.~\eqref{metric_transf}, in unitary gauge one obtains
\be
\hat K^\mu_{ \ \nu} = f^{-1/2} \left( K^\mu_{\  \nu} + \frac{\sqrt{- g^{00}}}2  \frac{\dot f}{f} h^\mu_{\ \nu}  \right)\;,
\ee
so that
\be
\delta \hat K^\mu_{ \ \nu} = f^{-1/2} \left( \delta { K}^\mu_{ \ \nu} - \frac{1}{4} \frac{ \delta  g^{00}}{\sqrt{- g^{00}}} \frac{\dot f}{f}  h^\mu_{\ \nu} \right) \;,
\ee 
where  $1/\sqrt{- g^{00}}$  can be expanded in polynomials of  $\delta  g^{00}$. Including in the conformal transformation the two last terms containing $\delta K$, the second line of eq.~\eqref{EFTDE_E2} becomes
\be
\begin{split}
d^4 x\sqrt{- g} & \left[  \left(\frac{f^2 M_2^4}{2} + \frac{3 f^{1/2} \bar m_1^3 \dot f}{8}  - \frac{9 \bar M_2^2 \dot f^2 }{32 f } \right) (\delta  g^{00})^2 \right. \\
&  +  \left( \frac{  f^2 M_3^4}{3!} + \frac{3 f^{1/2} \bar m_1^3 \dot f}{16} - \frac{9 \bar M_2^2  \dot f^2}{64 f} \right) (\delta  g^{00})^3 + \ldots \\ 
& \left.- \left( \frac{f^{3/2} \bar m_1^3}{2} - \frac{3 \bar M^2_2 \dot f}{4} \right)\delta  g^{00} \delta  K  - \frac{f \bar M_2^2}{2} \delta K^2+ \ldots  \right]\;. \label{ac3}
\end{split}
\ee
From the above, one can now easily recognize all the different pieces in the J-frame action~\eqref{ac} that we re-derived in the last two subsections from the E-frame perspective. The inclusion of other terms in the transformation is straightforward as well as the inverse transformation from Jordan to Einstein frame. The latter is done explicitly in App.~\ref{jortoein}.

\section{Concluding remarks}

We proposed a unifying description for dark energy and modified gravity. The aim is to recover all single-field models in the regime where cosmological perturbation theory is applicable and the background scalar is monotonic in time around the time-scales that are relevant for observations.
Our DE action~\eqref{ac} is organized as an expansion of increasing order in number of perturbations. This is made natural by the use of unitary gauge. In this gauge the dynamics of both the gravitational and scalar sectors is encoded in the degrees of freedom of the metric. Therefore, our DE action is written directly in terms of  the perturbations of the metric field around a FRW solution.  Only three operators, $f(t)$, $c(t)$ and  $\Lambda(t)$, contribute to linear order and are thus fixed by the background evolution. All the others start at least quadratic in the perturbations.  Each operator is responsible for distinctive dynamical features. 

In Sec.~\ref{sec_mixing}  we initiated a systematic study of such features. We considered a restricted number of notable operators in the limit of high energy. This is sufficient to study the ghost and gradient stability and to compute the speed of sound of the associated scalar fluctuation---we leave a more detailed analysis of the Jeans instability and of the clustering behavior to future work. Note that all those features can be studied once and for all in our formalism, and 
then any new  DE model is analyzed straightforwardly, once it is ``distilled" into our basic set of operators. For example, once the coefficients $\dot f$, $c$, $M_2^4$ and $\bar m_1^3$ have been worked out for a given DE model, its stability and speed of sound will be simply given by eqs.~\eqref{pai} and~\eqref{cies}. 

Our formalism also allows to address general issues in an efficient and easy way. For instance, there has been interest in understanding whether one can have a sensible theory with violations of the null energy condition or, in other words, a ``super-accelerating'' equation of state $w<-1$. For minimally coupled models this has been addressed  in~\cite{EFT1,Creminelli:2008wc}, and we can review the argument here by considering the limit $\dot f = 0$ of our theory. By the background equation~\eqref{c2},  $w<-1$ implies $c<0$. We can ask whether the Lagrangian for the fluctuations can still have the ``right" signs and the theory be stable. By looking at eq.~\eqref{pai} it is immediate to realize that ghost stability---related to the sign of the $\dot \pi^2$ term---is guaranteed for high enough values of $M_2^4$, $\bar m_1^3$. In turn, gradient stability will set an upper limit on the ratio $\bar m_1^3/(\MM^2 H)$ (see, e.g.,~\cite{Deffayet:2010qz} for an explicit model displaying this property). The present framework allows to generalize this argument to non-minimally coupled models ($\dot f \neq 0$), which are also known to provide an effective equation of state $w<-1$~\cite{Das:2005yj}. More generally, the question of the stability can be  immediately addressed from eqs.~\eqref{c2} and~\eqref{pai}, where it reduces to an algebraic problem. Without the need of tedious model building, a large variety of scenarios can be explored simply by varying the relative contributions of $c$, $M_2^4$ $\dot f$ and $\bar m_1^3$.

Our description looks handy also for an efficient comparison against the data. The main reasons are the formulation in terms of an action principle and the fact that  effects of quadratic and higher-order operators are disentangled from the background evolution. A more explicit link with observations is left to future work. 
In particular, it will be important to extend the dynamics beyond the strict high-energy limit considered here, in which case matter species acquire an important role. The aim is to match the operators of our description to observation of the large-scale structures, both in the linear and non-linear regime, and including higher-order statistics.

\subsection*{Acknowledgments} We wish to thank Paolo Creminelli, Cedric Deffayet, J\'er\^ome Gleyzes, Lam Hui, Justin Khoury, David Langlois, Alberto Nicolis and Ignacy Sawicki for discussions. G.G. and F.V. acknowledge the Paris Center for Cosmological Physics for kind hospitality.

\appendix

\section{Curvature invariants}
\label{sec_invariants} 

Here we explain how to include in the unitary-gauge action general curvature invariant  terms by considering $R_{\mu \nu}R^{\mu \nu}$ as an example. We follow closely (and refine here and there) the method outlined in App.~B of Ref.~\cite{EFT2}.

All curvature invariants are made with the Riemann tensor $R_{\mu \nu \rho \sigma}$. In $S_{DE}^{(2)}$ we want to include terms that start explicitly quadratic or higher in the perturbations, which means polynomials of at least second order in $\delta R_{\mu \nu \rho \sigma} \!= R_{\mu \nu \rho \sigma} \!- R_{\mu \nu \rho \sigma}^{(0)}$, $\delta R_{\mu \nu} \!= R_{\mu \nu } \!- R_{\mu \nu }^{(0)}$ or $\delta R = R - R^{(0)}$. The trick used in~\cite{EFT2} is to exploit the high degree of symmetry of FRW in order to write the curvature perturbations  in such a way that they are still covariant tensors. We thus \emph{define}\footnote{Note that this is not exactly eq.~(77) of Ref.~\cite{EFT2}:  we have corrected few typos and chosen to use $n_\mu$ instead of $\delta_\mu^0$ in the definition.} $R_{\mu \nu \rho \sigma}^{(0)}$
as
\begin{align} \label{R(0)}
R^{(0)}_{\mu \nu \rho \sigma} &= 2 \left(H^2 + \frac{k}{a^2}\right) h_{\mu [ \rho} h_{\sigma ] \nu} - \frac14 \left[ (H^2 + \dot H) h_{\mu \sigma} n_\nu n_\rho + {\rm perm.}\right]\, ,
\end{align}
where the brackets denote the antisymmetric part on the indicated indices. 
Analogously, we define $K^{(0)}_{\mu \nu} = H h_{\mu \nu}$, which is relevant for what follows. Note that we have defined a background quantity by using the ``full" tensor $g_{\mu \nu}$ and vector $n_\mu$.  From~\eqref{R(0)}, 
\begin{align} 
R^{(0)}_{\mu \nu} & = 2 \left(H^2 + \frac{k}{a^2}\right) h_{\mu \nu} - (H^2 + \dot H) (3 n_\mu n_\nu - h_{\mu \nu})\;,\label{R0}\\
R^{(0)} & = 6 \left(2 H^2 + \dot H +\frac{k}{a^2}\right) \label{R0munu} 
\end{align}
follow.
Of the three dimension-four operators $R^2$, $R_{\mu\nu}R^{\mu\nu}$ and $R_{\mu\nu\rho\sigma}R^{\mu\nu\rho\sigma}$, the latter can be substituted in favor of the other two and of the Gauss-Bonnet topologically invariant.
\begin{equation}
{\cal G} = R^2 - 4 R_{\mu\nu}R^{\mu\nu} + R_{\mu\nu\rho\sigma}R^{\mu\nu\rho\sigma} \;.
\end{equation}
Another choice is to use quadratic combinations of the Weyl tensor $C_{\mu \nu \rho \sigma}$ \cite{wein-eft} defined in eq.~\eqref{Weyl}, which in our formalism is particularly handy because its background part vanishes on a FRW background.

As $R^2$ can be expanded straightforwardly, let's consider $R_{\mu\nu}R^{\mu\nu}$ in some detail. Following~\cite{EFT2} we write
\begin{equation} \label{R^2}
R_{\mu\nu}R^{\mu\nu} = \delta R_{\mu\nu} \delta R^{\mu\nu} + 2 R_{\mu\nu}^{(0)} R^{\mu\nu} - R_{\mu\nu}^{(0)} R^{(0)\, \mu\nu} \, .
\end{equation}
The last term gives 
\begin{equation}
R_{\mu\nu}^{(0)} R^{(0)\, \mu\nu} = 12 \left(3 H^4 + \dot H^2 + 3 H^2 \frac{k}{a^2} + 3 H^2 \dot H + \dot H \frac{k}{a^2} + \frac{k^2}{a^4}\right) \;,
\end{equation}
and contributes to $\Lambda(t)$. 
For the second term on the right-hand side of~\eqref{R^2} we find
\begin{equation}
R_{\mu\nu}^{(0)} R^{\mu\nu} = 2 \left(\frac{k}{a^2} - \dot H\right) n_\mu n_\nu R^{\mu \nu} + (H^2 + \dot H) R \;. 
\end{equation}
In turn (see e.g.~\cite{Wald:1984rg}),
\begin{equation} 
n_\mu n_\nu R^{\mu \nu}  = K^2 - K_{\mu \nu} K^{\mu \nu} - \nabla_\mu(n^\mu \nabla_\nu n^\nu) + \nabla_\nu(n^\mu \nabla_\mu n^\nu)\;,
\end{equation} 
The last two terms are dealt with in App.~B of~\cite{EFT2}: the first of the two gives something proportional to $K \sqrt{- g^{00}}$, which can be expanded in powers of $\delta g^{00}$. The last vanishes in unitary gauge. In summary, we find that an operator of the form $A(\phi) R_{\mu\nu}R^{\mu\nu} \rightarrow A(t) R_{\mu\nu}R^{\mu\nu} $ contributes to the terms in~\eqref{ac} as follows (to simplify the formulas we set $k=0$ in the following):
\be
\begin{split}
f(t) & = 2 A(t)\left(3 H^2  + \dot H \right) \;, \\
\Lambda (t) & =  4 A(t) \left(9 H^4 + 5 \dot H^2 + 3 H^2 \dot H \right) + 2 H (A \dot H)\dot\  + 2 (A \dot H) \ddot\ \;,\\
c(t) &  = - 14 H (A \dot H) \dot\ - 8 A \dot H^2 - 2 (A \dot H)\ddot\ \;,\\
M^4_2 & =  14 H (A \dot H) \dot\ + 8 A \dot H^2 + 2 (A \dot H)\ddot\ \;,\\
\bar m^3_i & = - (A \dot H) \dot\ \;, \quad \bar M_2^2 =  -\bar M_3^2 = 2 (A \dot H) \;, \quad \lambda_2 = 2 A \;.
\end{split}
\ee

\section{Newtonian gauge}


After re-introducing $\pi$ with the Stueckeberg trick, we are free again to choose the most appropriate gauge. 
 The Newtonian gauge  is particularly suited for late-time cosmology. In this gauge the metric is in the form (we assume hereafter a spatially flat FRW background)
\begin{equation} \label{newtonian2} 
ds^2 = -e^{2\F}dt^2 + a^2 e^{-2 \PP} \delta_{i j} dx^i dx^j\, .
\end{equation}
It is not too long to calculate exactly the Christoffel symbols, 
\begin{align}
\Gamma_{00}^0 & = \dot \F \;, &\
\Gamma_{0i}^0  &= \partial_i \F\;, & \
\Gamma_{ij}^0  &= a^2 \delta_{ij} \, e^{- 2 \PP - 2 \F} \, (H - \dot \PP) \;, \label{Christ1}\\
\Gamma_{00}^i & = a^{-2}\, e^{ 2 \PP + 2 \F} \, \partial_i \PP \;, &\
\Gamma_{0j}^i  & = \delta_{ij} (H - \dot \PP) \;, &\
\Gamma_{jk}^i  &= \delta_{jk} \partial_i \PP - \delta_{ij} \partial_k \PP - \delta_{ik} \partial_j \PP\;.\label{Christ2}
\end{align}

With the aid of the above expressions we can calculate the extrinsic curvature defined in~\eqref{definition_ext} as $K_{\mu \nu} \equiv h^\sigma_\mu \nabla_\sigma n_\nu$. By~\eqref{nmu}, we have that in Newtonian gauge $n_\mu = - \delta_{\mu 0} e^\Phi$. Therefore,
\begin{equation}
\nabla_\rho n_\nu = - \delta_{\nu 0} e^{\Phi} \partial_\rho \Phi  + \Gamma_{\rho \nu}^0 e^\Phi \;.
\end{equation}
On the other hand, the projector $h_{\mu \nu}$ gives, for the space-space components  $h_{ij}=\delta_{i j} a^2 e^{-2 \Psi}$ and zero otherwise. 
By using the expressions of the Christoffel symbols, eqs.~\eqref{Christ1}--\eqref{Christ2}, we finally get that the only non zero components of $K_{\mu \nu}$ are
\begin{equation}
K_{i j} = a^2 \delta_{ij} e^{-2 \Psi - \Phi} (H - \dot \Psi) \simeq a^2 \delta_{ij} (H - \dot \Psi - H \Phi - 2 H \Psi) \;.
\end{equation}
We also get
\begin{equation} \label{deltaK}
\delta K = - 3(\dot \Psi + H \Phi) \;.
\end{equation}

\subsection{Gravitational action}
\label{app_formulas}

The Ricci scalar can also be written exactly in a rather compact form:
\be
\begin{split}
R \ = \ &  \, 6\, e^{- 2 \Phi}(2 H^2 + \dot H - H \dot \Phi - 4 H \dot \Psi + \dot \F \dot \PP + 2 \dot \Psi^2 - \ddot \Psi) \\
& - 2 a^{-2} e^{2 \PP} \big[ \nabla^2 \F - 2 \nabla^2 \PP + (\vec \nabla \F)^2 - \vec \nabla \F \vec \nabla \PP + (\vec \nabla \PP)^2 \big] \, ,
\label{Ricci_New}
\end{split}
\ee
where $\vec \nabla$ is a 3-d gradient and $\nabla^2$ the Laplacian.

It is useful at this point to write the Einstein-Hilbert term at quadratic order. After few integrations by parts we obtain
\begin{align} \label{EH}
\int \ \sqrt{-g}R \  \overset{\rm quadratic}{=}  \ \int&\ a\left[2 (\vec \nabla \Psi)^2 - 4 \vec \nabla \Phi \vec \nabla \Psi \right] - 6 a^3 \dot \Psi^2 \\ 
&- 3 a^3 \left[ 4 H\F \dot \Psi + H^2 (\Phi+ 3 \Psi)^2 - 6 \Psi^2 (3 H^2 + \dot H) \right]\, .\nonumber
\end{align}
Finally we can multiply the Einstein-Hilbert action for the $f(t)$ coefficient as in~\eqref{ac}. By the Stueckelberg trick $f(t)$ produces a $\pi$ field, which therefore mixes with gravity, 
\begin{align}
\int \ &\sqrt{-g}f(t) R \ \rightarrow \ \int \ \sqrt{-g}\left[f(t) + \dot f(t) \pi + \frac12 \ddot f \pi^2 + \dots \right]R\\ \label{141}
 \overset{\rm quadratic}{=} & \ \int \ a f \left[ 2 (\vec\nabla \Psi)^2 - 4 \vec \nabla \Phi \vec \nabla \Psi \right] + 2 a \dot f \pi \left(2 \nabla^2 \Psi - \nabla^2 \Phi\right) - 6 a^3 (f \dot \Psi^2 + \dot f \pi \ddot \Psi) \\
 & - 3 a^3 f \left[2 \F \dot \Psi \left( 2 H + \dot f/f\right)  + \, (\Phi+ 3 \Psi)^2 \left(H^2 + H\dot f/f\right) \right. \\
& \left.  - 3 \Psi^2 \left(6 H^2 + 2 \dot H + 5 H \dot f/f + \ddot f/f - \dot f^2/f^2\right)  \right] \\
&  - 6 a^3 \dot f \pi \left[H \dot\Phi + 4 H \dot \Psi + (2 H^2 + \dot H)(\Phi + 3 \Psi)\right] + a^3 \ddot f \pi^2 (6 H^2 + 3 \dot H)\, .
\end{align}
The first line,~\eqref{141}, is the most important because it contains the kinetic terms. This is the piece of the action that survives in the Minkowski limit\footnote{In Einstein frame this would be called the ``decoupling limit". In the Jordan frame however, as apparent, $\pi$ and gravity are kinetically mixed and thus not decoupled by taking the high-energy limit.}. The latter is obtained by posing $H = \dot H = \dot f = 0$. 
The reason  why the $\dot f$-terms in line~\eqref{141} survive in the Minkowski limit is that kinetic term for $\pi$ is multiplied by the coefficient $c$ (see e.g. eq~\eqref{pai}), which means that the canonically normalized field is $\pi_c \sim \pi c^{1/2}$. In taking the Minkowski limit, the ratio $\dot f/c^{1/2}$ should be kept constant and therefore also the combination $\dot f \pi$. 
More explicitly, the Minkowski limit reads 
\begin{equation}
\int \ \sqrt{-g}f(t) R  \overset{\rm kinetic}{=}  \ \int \  2 f (\vec \nabla \Psi)^2 - 4 f \vec \nabla \Phi \vec \nabla \Psi + 2 \dot f \pi \left(2 \nabla^2 \Psi - \nabla^2 \Phi\right) - 6 ( f \dot \Psi^2 + \dot f \pi \ddot \Psi)\, .
\end{equation}

\subsection{Linearized Einstein equations}
\label{app_linear}
We want to derive the linearized Einstein equations in Newtonian gauge~\eqref{newtonian2} for action~\eqref{acsecondein}.
In this gauge the linearized Einstein tensor components read
\begin{align}
G_{00}=&\ 3H^{2}-6H\dot\Psi+2a^{-2}\nablav^{2}\Psi\, , \nonumber\\
G_{ij}=& \ - a^{2} \big[3H^{2} +2\dot H \big] \delta_{i j} \\ &+a^2 \big[2H(\dot\Phi+3\dot\Psi)+2\ddot\Psi+2(\Phi+\Psi)(3H^{2}+2\dot H) \big] \delta_{i j}+ \, \nablav^{2}(\Phi- \Psi) \delta_{ij} + \partial_i \partial_{j}(\Psi - \Phi)\, , \nonumber \\
G_{0i}=&\ 2\partial_{i}(\dot \Psi + H \Phi)\, . \nonumber
\end{align}
Of course, since we want to study perturbations in Newtonian gauge, we have to reintroduce the scalar degree of freedom $\pi$ \emph{via} the Stueckelberg trick. It is handy to do so at the level of the action~\eqref{acsecondein}.
The contribution to the Einstein equations of the first three terms is given in eq.~\eqref{einsteine}.
On the other hand, the operator ${M_{2}^{4}}$ affects only the $(00)$ component yielding
$4M_{2}^{4}(\Phi-\dot \pi)$.

In order to find the contribution from the operator $ {\bar m_{1}^{3}}$ we first reintroduce the $\pi$ in the action,
\begin{equation}
\delta g^{00}\delta K\rightarrow(g^{00}+1+2g^{0\mu}\partial_{\mu}\pi)(K-3H+3\dot H \pi-a^{-2}\nablav^{2}\pi) \;,
\end{equation}
and we write the variation of $K$ as
\begin{equation}
\delta K\equiv \delta (\nabla_{\alpha}n^{\alpha})= \nabla_{\alpha}(n_{\beta}\delta g^{\alpha\beta})+\frac{1}{2}\frac{1}{-g^{00}}\nabla_{\alpha} (n^{\alpha}\delta g^{00})-\frac{1}{2}g_{\alpha\beta}n^{\sigma}\nabla_{\sigma}(\delta g^{\alpha\beta})\;.
\end{equation}
Then, upon integration by parts, we obtain
\begin{align} 
\frac{1}{\sqrt{-g}} \frac{\delta (\sqrt{-g} \bar m_1^3 \delta g^{00} \delta K)}{\delta g^{\mu \nu}} \to & \ \bar m_{1}^{3}\delta_{\mu}^{0}\delta_{\nu}^{0}(K-3H+3\dot H \pi-a^{-2}\nablav^{2}\pi)+\frac{\bar m_{1}^{3}}{2}(g^{00}+1-2\dot\pi)K g_{\mu\nu}\nonumber\\
&-\frac{1}{2} \nabla_{\alpha}(\bar m_{1}^{3}(g^{00}+1-2\dot\pi))\left( n^{\alpha}n_{\mu}n_{\nu}+2n_{\nu}\delta^{\alpha}_{\mu}-n^{\alpha}g_{\mu\nu}\right) \;,
\end{align}
where, at first order in metric perturbations,
$K-3H = -3(\dot\Psi+H\Phi)$, $g^{00}+1=2\Phi $
and the $n_{\alpha}$'s appearing explicitly only need to be computed at zero order in the perturbations.

Taking into account  these last two terms that we have discussed and assuming $\bar m_1^3$ constant, the linearized Einstein equations read
\begin{list}{\labelitemi}{\leftmargin=.1em}
\item $(00)$ component:
\begin{align}
&\MM^2\left[2 f ( a^{-2}\nablav^{2}\Psi  -3H\dot\Psi ) + 3 \dot f(-\dot\Psi + H^{2}\pi + H \dot \pi - a^{-2} \nablav^{2}\pi/3 )+3 \ddot f H \pi \right]\nonumber \quad \quad\quad \quad\quad \quad\quad \quad \quad \\
&-2c\dot\pi-(\dot c+\dot\Lambda)\pi-2\Lambda\Phi+4M_{2}^{4}(\Phi-\dot \pi)\\
&+\bar m_{1}^{3}\left[3(\dot\Psi+H\Phi)-3\dot H\pi+a^{-2}\nablav^{2}\pi+3H (\Phi-\dot \pi)\right]= \delta T_{00}\, .\nonumber
\end{align}
By using the equations of motion \eqref{c2}, \eqref{L2}, \eqref{frie1} and $\delta T_{00} = \delta \rho _m + 2 \rho_m \Phi$ we get
\begin{align}
&\MM^2\left[2 f \big( a^{-2}\nablav^{2}\Psi  -3H(\dot\Psi + H \Phi) \big) - \dot f \big(3(\dot\Psi+ H\Phi )+ 3 \dot H \pi+a^{-2}\nablav^{2}\pi +2 H (\Phi-\dot \pi) \big) \right. \nonumber\qquad \qquad \\  &\left.- \ddot f (\Phi - \dot \pi)\right]
+ (\rho_D + p_D) (\Phi - \dot \pi + 3 H \pi) + 4M_{2}^{4}(\Phi-\dot \pi) \\
&+\bar m_{1}^{3}\left[3(\dot\Psi+H\Phi)-3\dot H\pi+a^{-2}\nablav^{2}\pi+3H (\Phi-\dot \pi)\right]= \delta \rho_m\, . \nonumber
\end{align}
\item $(i j)$-trace components:
\begin{align}
&\MM^2\left\{ 2 f\left[\ddot\Psi +H \dot\Phi+3H \dot\Psi+ (3H^{2}+ 2\dot H) (\Phi+\Psi) +  \nablav^2(\Phi - \Psi)/(3 a^2)\right] \right.\nonumber\\
&\left. + \dot f\left[\dot \Phi + 2 \dot \Psi + 4 H(\Phi+\Psi)  - (3H^{2}+2\dot H)\pi  + 2 \nablav^2 \pi/(3 a^2)\right] - 2 H(\dot f \pi)\dot\ - (\dot f \pi)\ddot\ + 2 \ddot f (\Phi + \Psi) \right\}\nonumber\\
&+2 c(\Phi-\dot\pi)-2\Psi (\Lambda-c)+ (\dot\Lambda-\dot c)\pi - \bar m_{1}^{3} [ \dot\Phi-\ddot\pi +3H(\Phi-\dot\pi)]=  \delta T^k_{\ k}/(3 a^2) \;.
\end{align}
Again, by use of the equations of motion, 
\begin{align}
&\MM^2\left\{ 2 f\left[\ddot\Psi +H \dot\Phi+3H \dot\Psi+ (3H^{2}+ 2\dot H) \Phi  +  \nablav^2(\Phi - \Psi)/(3 a^2)\right] \right.\nonumber \qquad \qquad \qquad \qquad \\
&\left. + \dot f\left[  2 (\dot \Psi  + H \Phi) + (\dot \Phi -\ddot \pi)  + 3 H (\Phi  -  \dot \pi) - 3H^{2}\pi  + 2 \nablav^2 \pi/(3 a^2)\right] +  \ddot f (\Phi - \dot \pi) \right\}\\
&- \dot p_D \pi  + (\rho_D + p_D) (\Phi- \dot \pi) - \bar m_{1}^{3} [ \dot\Phi-\ddot\pi +3H(\Phi-\dot\pi)] = \delta p_m \nonumber \;.
\end{align}
\item $(i j)$-traceless components:
\begin{equation}
\MM^2 \big(\partial_{i}\partial_{j} - \frac13 \delta_{ij} \nablav^2 \big) \left[f(\Psi-\Phi)- \dot f \pi\right]= \delta T_{ij}  - \frac13 \delta_{ij} \delta T^k_{\ k}\;. \label{eq_traceless}
\end{equation}
\item $(0i)$ components:
\begin{equation}
\MM^2\partial_{i}\left[ 2f(\dot \Psi+H\Phi)+(\ddot f-H\dot f) \pi+\dot f(\Phi-\dot \pi) \right]- 2(\rho_{D}+p_{D})\partial_{i}\pi - 2\bar m_{1}^{3}\partial_{i}(\Phi-\dot\pi)=\delta T_{i0} \;.
\end{equation}

\end{list}

\section{Field redefinitions of the metric}
\label{app_3}
\subsection{Perturbative field redefinition in the Einstein frame}
\label{app:field_red} 

This subsection of App.~\ref{app_3} and Sec.~\ref{sec_general} are the only places in the paper where Einstein-frame quantities {\em are not} denoted with a hat nor with the subscript $E$ and Jordan-frame quantities explicitly carry a subscript $J$.
Here we show that  the action
\be
\label{Eac1}
S=   \int d^4x \sqrt{-g} \left[ \frac{\MM^2 R }{2}  - c(t) g^{00} -  \Lambda(t) +  F^{(2)} (\delta g^{00}, \delta K_{\mu \nu}, \ldots ; t ) \right] + S_{m} \left[ g_{J }^{\mu \nu}, \psi_{m} \right] \;,
\ee
can be rewritten as 
\be
\label{Eac2}
S=   \int d^4x \sqrt{-g} \left[ \frac{\MM^2 R }{2}  - \tilde c(t) g^{00} -  \Lambda(t) + \tilde F^{(2)} (\delta g^{00}, \delta K_{\mu \nu}, \ldots ; t ) \right] + S_{m} \left[  f(t) g^{\mu \nu}, \psi_{i} \right] \;.
\ee
Above, $g_J^{\mu \nu}$ is defined as in eq.~\eqref{JE}, i.e.,
\be
\begin{split}
\label{JE2}
g_J^{\mu \nu} = & \ f( t)  g^{\mu \nu}  +   g^{\mu \nu} \left[ \beta_1 ( t) \delta  g^{00} + \frac12 \beta_2 ( t) (\delta  g^{00})^2 + \dots  \right] +  \delta  g^{0 \mu} \delta  g^{0\nu} \left[\gamma_0( t) + \gamma_1 ( t) \delta  g^{00}  + \dots  \right]  \\
&+ \frac12 m_1^{-1}( t)  g^{\mu \nu} \delta g^{00}   K
+ \frac12 m_{2}^{-2}( t) \delta  K^{\mu \rho} \delta  K_{\rho}^{\ \nu} + \ldots\\[2mm]
\equiv & \ f( t)  g^{\mu \nu}  + \delta F^{\mu \nu}\;. 
\end{split}
\ee

Consider now the field-redefinition of the metric 
\begin{equation} \label{fieldred}
g^{\mu \nu} \to \tilde g^{\mu \nu} =  g^{\mu \nu} +f^{-1} \delta F^{\mu \nu}. 
\end{equation}
In terms of the new field $\tilde g^{\mu \nu}$, the matter action has already the form that we want,
\be
S_{m} \left[  g_{J }^{\mu \nu}, \psi_{i} \right] \ = \  S_{m} \left[  f(t) \tilde g^{\mu \nu}, \psi_{i} \right]\;.
\ee
Let us now concentrate on the DE part of the action and in particular on the first three terms inside the square brackets in eq.~\eqref{Eac1}. To see the effects of a perturbative field redefinition consider the linear term in the functional Taylor expansion, 
\begin{equation} 
S[\tilde g^{\mu \nu} - f^{-1} \delta F^{\mu \nu}] \, \simeq \, S[\tilde g^{\mu \nu}] - \frac{\delta S}{\delta g^{\mu \nu}} f^{-1} \delta F^{\mu \nu}  = \, S[\tilde g^{\mu \nu}] - \int \frac{\sqrt{-\tilde g}}{2}  f^{-1} (t) \bar T(t) \beta_1(t) \tilde g^{00} + \dots\, ,
\end{equation}
where the ellipses denote second-order terms or higher. For the second equality we have used the background Einstein's equation, eq.~\eqref{Ein_Ein2}, and for $\delta F^{\mu \nu}$ we have replace the second term of the Jordan frame metric, eq.~\eqref{JE2}. Thus, this Lagrangian transforms in itself plus a linear term in $g^{00}$ proportional to the parameter $\beta_1(t)$, which redefines the function $c(t)$,
\be
c(t) \to \tilde c(t) = c(t) + \frac12  \bar T(t) \frac{\beta_1(t)}{f(t)}\;,
\ee
plus second-order or higher terms which do not involve the matter energy-momentum tensor.
The higher-order terms $F^{(2)} $ in eq.~\eqref{Eac1} are clearly affected by the field redefinition. However, their structure remains unaltered, as they can always be rewritten as polynomials of the geometrical invariant objects $\delta g^{00}, \delta K_{\mu \nu}$, etc., following \cite{EFT1}. In conclusion, the action after the field redefinition is eq.~\eqref{Eac2}.

\subsection{From Jordan to Einstein} 
\label{jortoein}

In this appendix Einstein frame quantities are denoted with an index $E$ or with a hat.  
In order to study the stability of our DE theory we consider action~\eqref{ac} keeping terms up to second order in the perturbations, 
\begin{equation} 
\begin{split}
S_{DE}=  &\int d^4x \sqrt{-g} \left[ \frac{\MM^2}{2} f(t) R  - c(t) g^{00} -  \Lambda(t) \right. \\ 
&\left. + \frac{M_2^4}{2} (\delta g^{00})^2  - \frac{\bar m_1^3}{2} \delta g^{00} \delta K - \frac{\bar M_2^2}{2} \delta K^2 - \frac{\bar M_3^2}{2} \delta K_{\mu}^{\ \nu} \delta K_{\ \nu}^\mu + \ldots \right]\;. \label{acsecond}
\end{split}
\ee
In order to switch to Einstein frame it will be useful to review some standard ``vocabulary" of conformal transformations,
\begin{align}
g_{\mu \nu}  & = f^{-1} \hg_{\mu \nu} \label{comps_trans}\\
 \sqrt{- g} & = f^{-2} \sqrt{- \hg}  \;, \\
R &= f \left(R_E + 3  \hat \square \ln f - \frac32  \hg^{\mu \nu}   \partial_\mu \ln f   \partial_\nu \ln f \right) \;,
\end{align}
from which we get
\begin{equation}
\delta K_\mu^\nu = f^{1/2}\left(\delta \hat K_\mu ^\nu + \frac{1}{4}\frac{\delta \hg^{00}}{\sqrt{-\hg^{00}}}\frac{\dot f}{f} h_\mu^\nu \right) \;, \qquad 
\delta K = f^{1/2}\left(\delta \hat K + \frac{3}{4}\frac{\delta \hg^{00}}{\sqrt{-\hg^{00}}}\frac{1}{f}\frac{\dot f}{f}\right)\, ,
\end{equation}
where $h$ is the metric projected on $t = const.$ surfaces. The problem with $\hat g^{00}$ so defined is that its background value is $-f^{-1}$ instead of $-1$. This is related to the fact that by conformally transforming the metric components (eq.~\eqref{comps_trans}) without changing coordinates one looses the gauge choice $g_{00} = g^{00} = -1$. While moving to the Einstein frame it is thus useful to redefine the time coordinate $dt_E = f^{1/2} dt$ in order to have the metric in the usual form. This was done systematically in Sec.~\ref{sec_etoj}. After such a time redefinition we find 
\begin{equation} 
\begin{split}
S_{DE}=  &\int d^4 x_E \sqrt{-\hat g} \left[ \frac{\MM^2}{2} R_E  - c_E \hat g^{00} -  \Lambda_E \right. \\ 
&\left. + \frac{\hat{M}_2^4}{2} (\delta \hat g^{00})^2  - \frac{\hat {\bar m}_1^3}{2} \delta \hat g^{00} \delta \hat K - \frac{\hat {\bar M}_2^2}{2} \delta \hat K^2 - \frac{\hat {\bar M}_3^2}{2} \delta \hat K_{\mu}^{\ \nu} \hat \delta K_{\ \nu}^\mu + \ldots \right]\;, 
\end{split}
\ee
with
\begin{align}
f^2 c_E &= c + \frac34 \MM^2 \frac{\dot f^2}{f} \; \qquad f^2 \Lambda_E =  \Lambda \label{celambda}\\
f^2 \hat M_2^4 &= M_2^4 - \frac34 \bar m_1^3\, (\dot f/ f) - \frac{9}{16} \bar M_2^2 \, (\dot f/ f)^2 - \frac{3}{16} \bar M_3^2 \, (\dot f/ f)^2  \;,\label{emmedue}\\
f^{3/2} \hat{\bar m}_1^3 &=  {\bar m}_1^3 + \frac32 \bar M_2^2 \, (\dot f/ f)+ \frac12 \bar M_3^2 \, (\dot f/ f) \;, \qquad 
f \hat{\bar M}_2^2 = {\bar M}_2^2 \;, \qquad f \hat{\bar M}_3^2 = {\bar M}_3^2\, ,
\end{align} 
where, on the right-hand side, a dot denotes the derivative with respect to J-frame time.


\footnotesize
\parskip 0pt

\begin{thebibliography}{99}

\bibitem{euclid1} http://www.euclid-ec.org/

\bibitem{euclid2} 
  R.~Laureijs, J.~Amiaux, S.~Arduini, J.~-L.~Augueres, J.~Brinchmann, R.~Cole, M.~Cropper and C.~Dabin {\it et al.},
  ``Euclid Definition Study Report,''
  arXiv:1110.3193 [astro-ph.CO].

\bibitem{bigboss} 
  D.~J.~Schlegel, C.~Bebek, H.~Heetderks, S.~Ho, M.~Lampton, M.~Levi, N.~Mostek and N.~Padmanabhan {\it et al.},
  ``BigBOSS: The Ground-Based Stage IV Dark Energy Experiment,''
  arXiv:0904.0468 [astro-ph.CO].

\bibitem{Ishak:2005zs} 
  M.~Ishak, A.~Upadhye and D.~N.~Spergel,
  ``Probing cosmic acceleration beyond the equation of state: Distinguishing between dark energy and modified gravity models,''
  Phys.\ Rev.\ D {\bf 74}, 043513 (2006)
  [astro-ph/0507184].

\bibitem{costas} 
  T.~Clifton, P.~G.~Ferreira, A.~Padilla and C.~Skordis,
  ``Modified Gravity and Cosmology,''
  Phys.\ Rept.\  {\bf 513}, 1 (2012)
  [arXiv:1106.2476 [astro-ph.CO]].

 \bibitem{lucashin} 
  L.~Amendola and S.~Tsujikawa,
  ``Dark Energy: Theory and Observations,''
  Cambridge U. P. (2011) 506~p

\bibitem{post-fried1} 
  M.~Tegmark,
  ``Measuring the metric: A Parametrized postFriedmanian approach to the cosmic dark energy problem,''
  Phys.\ Rev.\ D {\bf 66}, 103507 (2002)
  [astro-ph/0101354].

\bibitem{post-fried2} 
  W.~Hu and I.~Sawicki,
  ``A Parameterized Post-Friedmann Framework for Modified Gravity,''
  Phys.\ Rev.\ D {\bf 76}, 104043 (2007)
  [arXiv:0708.1190 [astro-ph]].

\bibitem{Creminelli:2008wc} 
  P.~Creminelli, G.~D'Amico, J.~Norena and F.~Vernizzi,
  ``The Effective Theory of Quintessence: the w<-1 Side Unveiled,''
  JCAP {\bf 0902}, 018 (2009)
  [arXiv:0811.0827 [astro-ph]].

  \bibitem{PZW} 
  M.~Park, K.~M.~Zurek and S.~Watson,
  ``A Unified Approach to Cosmic Acceleration,''
  Phys.\ Rev.\ D {\bf 81}, 124008 (2010)
  [arXiv:1003.1722 [hep-th]].


\bibitem{Baker:2011jy} 
  T.~Baker, P.~G.~Ferreira, C.~Skordis and J.~Zuntz,
  ``Towards a fully consistent parameterization of modified gravity,''
  Phys.\ Rev.\ D {\bf 84}, 124018 (2011)
  [arXiv:1107.0491 [astro-ph.CO]].

  \bibitem{BF} 
  J.~K.~Bloomfield and E.~E.~Flanagan,
  ``A Class of Effective Field Theory Models of Cosmic Acceleration,''
  arXiv:1112.0303 [gr-qc].

\bibitem{Jimenez:2012jg} 
  R.~Jimenez, P.~Talavera, L.~Verde, M.~Moresco, A.~Cimatti and L.~Pozzetti,
  ``The effective Lagrangian of dark energy from observations,''
  JCAP {\bf 1203}, 014 (2012)
  [arXiv:1201.3608 [astro-ph.CO]].


\bibitem{Battye:2012eu} 
  R.~A.~Battye and J.~A.~Pearson,
  ``Effective action approach to cosmological perturbations in dark energy and modified gravity,''
  JCAP {\bf 1207}, 019 (2012)
  [arXiv:1203.0398 [hep-th]].

\bibitem{Baker:2012zs} 
  T.~Baker, P.~G.~Ferreira and C.~Skordis,
  ``The Parameterized Post-Friedmann Framework for Theories of Modified Gravity: Concepts, Formalism and Examples,''
  arXiv:1209.2117 [astro-ph.CO].

\bibitem{Sawicki:2012re} 
  I.~Sawicki, I.~D.~Saltas, L.~Amendola and M.~Kunz,
  ``Consistent perturbations in an imperfect fluid,''
  arXiv:1208.4855 [astro-ph.CO].

\bibitem{EFT1} 
  P.~Creminelli, M.~A.~Luty, A.~Nicolis and L.~Senatore,
  ``Starting the Universe: Stable Violation of the Null Energy Condition and Non-standard Cosmologies,''
  JHEP {\bf 0612}, 080 (2006)
  [hep-th/0606090].

\bibitem{EFT2} 
  C.~Cheung, P.~Creminelli, A.~L.~Fitzpatrick, J.~Kaplan and L.~Senatore,
  ``The Effective Field Theory of Inflation,''
  JHEP {\bf 0803}, 014 (2008)
  [arXiv:0709.0293 [hep-th]].

\bibitem{ghost}
  N.~Arkani-Hamed, H.~-C.~Cheng, M.~A.~Luty, S.~Mukohyama,
 ``Ghost condensation and a consistent infrared modification of gravity,''
  JHEP {\bf 0405}, 074 (2004).
  [hep-th/0312099].

\bibitem{multifield} 
  L.~Senatore and M.~Zaldarriaga,
  ``The Effective Field Theory of Multifield Inflation,''
  JHEP {\bf 1204}, 024 (2012)
  [arXiv:1009.2093 [hep-th]].
  
  \bibitem{dissipative} 
  D.~Lopez Nacir, R.~A.~Porto, L.~Senatore and M.~Zaldarriaga,
  ``Dissipative effects in the Effective Field Theory of Inflation,''
  JHEP {\bf 1201}, 075 (2012)
  [arXiv:1109.4192 [hep-th]].

\bibitem{Senatore:2009gt} 
  L.~Senatore, K.~M.~Smith and M.~Zaldarriaga,
  ``Non-Gaussianities in Single Field Inflation and their Optimal Limits from the WMAP 5-year Data,''
  JCAP {\bf 1001}, 028 (2010)
  [arXiv:0905.3746 [astro-ph.CO]].

\bibitem{Creminelli:2010qf} 
  P.~Creminelli, G.~D'Amico, M.~Musso, J.~Norena and E.~Trincherini,
  ``Galilean symmetry in the effective theory of inflation: new shapes of non-Gaussianity,''
  JCAP {\bf 1102}, 006 (2011)
  [arXiv:1011.3004 [hep-th]].
  
  \bibitem{daniel} 
  D.~Baumann and D.~Green,
  ``Equilateral Non-Gaussianity and New Physics on the Horizon,''
  JCAP {\bf 1109}, 014 (2011)
  [arXiv:1102.5343 [hep-th]].

\bibitem{lamnic} 
  L.~Hui and A.~Nicolis,
  ``An Equivalence principle for scalar forces,''
  Phys.\ Rev.\ Lett.\  {\bf 105}, 231101 (2010)
  [arXiv:1009.2520 [hep-th]].

\bibitem{picon} 
  C.~Armendariz-Picon and R.~Penco,
  ``Quantum Equivalence Principle Violations in Scalar-Tensor Theories,''
  Phys.\ Rev.\ D {\bf 85}, 044052 (2012)
  [arXiv:1108.6028 [hep-th]].

\bibitem{NP} 
  F.~Nitti and F.~Piazza,
  ``Scalar-tensor theories, trace anomalies and the QCD-frame,''
  [1202.2105 [hep-th]].

\bibitem{malda} 
  J.~M.~Maldacena,
  ``Non-Gaussian features of primordial fluctuations in single field inflationary models,''
  JHEP {\bf 0305}, 013 (2003)
  [astro-ph/0210603].

\bibitem{usep1} 
  F.~Piazza,
  ``The IR-Completion of Gravity: What happens at Hubble Scales?,''
  New J.\ Phys.\  {\bf 11}, 113050 (2009)
  [arXiv:0907.0765 [hep-th]].

\bibitem{usep2} 
  F.~Piazza,
  ``Infrared-modified Universe,''
  arXiv:1204.4099 [gr-qc].

  \bibitem{Amendola:1999er} 
  L.~Amendola,
  ``Coupled quintessence,''
  Phys.\ Rev.\ D {\bf 62}, 043511 (2000)
  [astro-ph/9908023].
  
  \bibitem{Gasperini:2001pc} 
  M.~Gasperini, F.~Piazza and G.~Veneziano,
  ``Quintessence as a runaway dilaton,''
  Phys.\ Rev.\ D {\bf 65}, 023508 (2002)
  [gr-qc/0108016].
  
  \bibitem{Comelli:2003cv} 
  D.~Comelli, M.~Pietroni and A.~Riotto,
  ``Dark energy and dark matter,''
  Phys.\ Lett.\ B {\bf 571}, 115 (2003)
  [hep-ph/0302080].

\bibitem{ArkaniHamed:2005gu} 
  N.~Arkani-Hamed, H.~-C.~Cheng, M.~A.~Luty, S.~Mukohyama and T.~Wiseman,
  ``Dynamics of gravity in a Higgs phase,''
  JHEP {\bf 0701}, 036 (2007)
  [hep-ph/0507120].

\bibitem{bean} 
  E.~-M.~Mueller, R.~Bean and S.~Watson,
  ``Cosmological Implications of the Effective Field Theory of Cosmic Acceleration,''
  arXiv:1209.2706 [astro-ph.CO].

\bibitem{wein-eft} 
  S.~Weinberg,
  ``Effective Field Theory for Inflation,''
  Phys.\ Rev.\ D {\bf 77}, 123541 (2008)
  [0804.4291 [hep-th]].



\bibitem{horndeski}
G.~W.~Horndeski, 
Int.\ J.\ Theor.\ Phys. {\bf 10}, 363 (1974).


\bibitem{Deffayet:2009mn} 
  C.~Deffayet, S.~Deser and G.~Esposito-Farese,
  ``Generalized Galileons: All scalar models whose curved background extensions maintain second-order field equations and stress-tensors,''
  Phys.\ Rev.\ D {\bf 80}, 064015 (2009)
  [arXiv:0906.1967 [gr-qc]].
  

\bibitem{fab-four} 
  C.~Charmousis, E.~J.~Copeland, A.~Padilla and P.~M.~Saffin,
  ``General second order scalar-tensor theory, self tuning, and the Fab Four,''
  Phys.\ Rev.\ Lett.\  {\bf 108}, 051101 (2012)
  [arXiv:1106.2000 [hep-th]].

  \bibitem{NRT}
A.~Nicolis, R.~Rattazzi and E.~Trincherini,
  ``The Galileon as a local modification of gravity,''
  Phys.\ Rev.\ D {\bf 79}, 064036 (2009)
  [arXiv:0811.2197 [hep-th]].
  
\bibitem{Deffayet:2009wt} 
  C.~Deffayet, G.~Esposito-Farese and A.~Vikman,
  Phys.\ Rev.\ D {\bf 79}, 084003 (2009)
  [arXiv:0901.1314 [hep-th]].

\bibitem{cliff} 
  C.~P.~Burgess,
  ``Quantum gravity in everyday life: General relativity as an effective field theory,''
  Living Rev.\ Rel.\  {\bf 7}, 5 (2004)
  [gr-qc/0311082].

\bibitem{BD} 
  C.~Brans and R.~H.~Dicke,
  ``Mach's principle and a relativistic theory of gravitation,''
  Phys.\ Rev.\  {\bf 124}, 925 (1961).


\bibitem{polarski} 
  B.~Boisseau, G.~Esposito-Farese, D.~Polarski and A.~A.~Starobinsky,
  ``Reconstruction of a scalar-tensor theory of gravity in an accelerating universe,''
  Phys.\ Rev.\ Lett.\  {\bf 85}, 2236 (2000)
  [gr-qc/0001066].


  \bibitem{h0} 
  W.~L.~Freedman {\it et al.}  [HST Collaboration],
  ``Final results from the Hubble Space Telescope key project to measure the Hubble constant,''
  Astrophys.\ J.\  {\bf 553}, 47 (2001)
  [astro-ph/0012376].

\bibitem{Damour:1992we} 
  T.~Damour and G.~Esposito-Farese,
  ``Tensor multiscalar theories of gravitation,''
  Class.\ Quant.\ Grav.\  {\bf 9}, 2093 (1992).


\bibitem{Luty:2003vm} 
  M.~A.~Luty, M.~Porrati and R.~Rattazzi,
  ``Strong interactions and stability in the DGP model,''
  JHEP {\bf 0309}, 029 (2003)
  [hep-th/0303116].
  
\bibitem{Nicolis:2004qq} 
  A.~Nicolis and R.~Rattazzi,
  ``Classical and quantum consistency of the DGP model,''
  JHEP {\bf 0406}, 059 (2004)
  [hep-th/0404159].
  

\bibitem{Deffayet:2010qz} 
  C.~Deffayet, O.~Pujolas, I.~Sawicki and A.~Vikman,
  ``Imperfect Dark Energy from Kinetic Gravity Braiding,''
  JCAP {\bf 1010}, 026 (2010)
  [arXiv:1008.0048 [hep-th]].


\bibitem{Das:2005yj} 
  S.~Das, P.~S.~Corasaniti and J.~Khoury,
  ``Super-acceleration as signature of dark sector interaction,''
  Phys.\ Rev.\ D {\bf 73}, 083509 (2006)
  [astro-ph/0510628].










\bibitem{ArmendarizPicon:2000dh} 
  C.~Armendariz-Picon, V.~F.~Mukhanov and P.~J.~Steinhardt,
  ``A Dynamical solution to the problem of a small cosmological constant and late time cosmic acceleration,''
  Phys.\ Rev.\ Lett.\  {\bf 85}, 4438 (2000)
  [astro-ph/0004134].
  
  
  \bibitem{woodard} 
  R.~P.~Woodard,
  ``Avoiding dark energy with 1/r modifications of gravity,''
  Lect.\ Notes Phys.\  {\bf 720}, 403 (2007)
  [astro-ph/0601672].

\bibitem{Starobinsky:1980te} 
  A.~A.~Starobinsky,
  ``A New Type of Isotropic Cosmological Models Without Singularity,''
  Phys.\ Lett.\ B {\bf 91}, 99 (1980).
  
  \bibitem{Capozziello:2003tk} 
  S.~Capozziello, S.~Carloni and A.~Troisi,
  ``Quintessence without scalar fields,''
  Recent Res.\ Dev.\ Astron.\ Astrophys.\  {\bf 1}, 625 (2003)
  [astro-ph/0303041].
  
\bibitem{Carroll:2003wy} 
  S.~M.~Carroll, V.~Duvvuri, M.~Trodden and M.~S.~Turner,
  ``Is cosmic speed - up due to new gravitational physics?,''
  Phys.\ Rev.\ D {\bf 70}, 043528 (2004)
  [astro-ph/0306438].

 \bibitem{Nojiri:2005jg} 
  S.~'i.~Nojiri and S.~D.~Odintsov,
  ``Modified Gauss-Bonnet theory as gravitational alternative for dark energy,''
  Phys.\ Lett.\ B {\bf 631}, 1 (2005)
  [hep-th/0508049].


 \bibitem{DeFelice:2010aj} 
  A.~De Felice and S.~Tsujikawa,
  ``f(R) theories,''
  Living Rev.\ Rel.\  {\bf 13}, 3 (2010)
  [arXiv:1002.4928 [gr-qc]].

  \bibitem{Chiba:2003ir} 
  T.~Chiba,
  ``1/R gravity and scalar - tensor gravity,''
  Phys.\ Lett.\ B {\bf 575}, 1 (2003)
  [astro-ph/0307338].
  
\bibitem{chame} 
  J.~Khoury and A.~Weltman,
  ``Chameleon cosmology,''
  Phys.\ Rev.\ D {\bf 69}, 044026 (2004)
  [astro-ph/0309411]; 
  ``Chameleon fields: Awaiting surprises for tests of gravity in space,''
  Phys.\ Rev.\ Lett.\  {\bf 93}, 171104 (2004)
  [astro-ph/0309300].
  
  \bibitem{Wald:1984rg} 
  R.~M.~Wald,
  ``General Relativity,''
  Chicago, Usa: Univ. Pr. ( 1984) 491p
  
\bibitem{Creminelli:2009mu} 
  P.~Creminelli, G.~D'Amico, J.~Norena, L.~Senatore and F.~Vernizzi,
  ``Spherical collapse in quintessence models with zero speed of sound,''
  JCAP {\bf 1003}, 027 (2010)
  [arXiv:0911.2701 [astro-ph.CO]].
  
\bibitem{Sefusatti:2011cm} 
  E.~Sefusatti and F.~Vernizzi,
  ``Cosmological structure formation with clustering quintessence,''
  JCAP {\bf 1103}, 047 (2011)
  [arXiv:1101.1026 [astro-ph.CO]].

\bibitem{D'Amico:2011pf} 
  G.~D'Amico and E.~Sefusatti,
  ``The nonlinear power spectrum in clustering quintessence cosmologies,''
  JCAP {\bf 1111}, 013 (2011)
  [arXiv:1106.0314 [astro-ph.CO]].

\bibitem{Lim:2010yk} 
  E.~A.~Lim, I.~Sawicki and A.~Vikman,
  JCAP {\bf 1005}, 012 (2010)
  [arXiv:1003.5751 [astro-ph.CO]].

\bibitem{Vainshtein:1972sx} 
  A.~I.~Vainshtein,
  ``To the problem of nonvanishing gravitation mass,''
  Phys.\ Lett.\ B {\bf 39}, 393 (1972).



\bibitem{Dvali:2000hr} 
  G.~R.~Dvali, G.~Gabadadze and M.~Porrati,
  ``4-D gravity on a brane in 5-D Minkowski space,''
  Phys.\ Lett.\ B {\bf 485}, 208 (2000)
  [hep-th/0005016].

\bibitem{Creminelli:2010ba} 
  P.~Creminelli, A.~Nicolis and E.~Trincherini,
  ``Galilean Genesis: An Alternative to inflation,''
  JCAP {\bf 1011}, 021 (2010)
  [arXiv:1007.0027 [hep-th]].

\bibitem{justin} 
  N.~Chow and J.~Khoury,
  ``Galileon Cosmology,''
  Phys.\ Rev.\ D {\bf 80}, 024037 (2009)
  [arXiv:0905.1325 [hep-th]].
  
\bibitem{Horava:2009uw} 
  P.~Horava,
  ``Quantum Gravity at a Lifshitz Point,''
  Phys.\ Rev.\ D {\bf 79}, 084008 (2009)
  [arXiv:0901.3775].
  
\bibitem{Blas:2009qj} 
  D.~Blas, O.~Pujolas and S.~Sibiryakov,
  ``Consistent Extension of Horava Gravity,''
  Phys.\ Rev.\ Lett.\  {\bf 104}, 181302 (2010)
  [arXiv:0909.3525 [hep-th]].
  
   
\bibitem{Blas:2010hb} 
  D.~Blas, O.~Pujolas and S.~Sibiryakov,
  ``Models of non-relativistic quantum gravity: The Good, the bad and the healthy,''
  JHEP {\bf 1104}, 018 (2011)
  [arXiv:1007.3503 [hep-th]].


\bibitem{Blas:2011en} 
  D.~Blas and S.~Sibiryakov,
  ``Technically natural dark energy from Lorentz breaking,''
  JCAP {\bf 1107}, 026 (2011)
  [arXiv:1104.3579 [hep-th]].

\bibitem{Blas:2012vn} 
  D.~Blas, M.~M.~Ivanov and S.~Sibiryakov,
  ``Testing Lorentz invariance of dark matter,''
  arXiv:1209.0464 [astro-ph.CO].



\bibitem{Creminelli:2012xb} 
  P.~Creminelli, J.~Norena, M.~Pena and M.~Simonovic,
  ``Khronon inflation,''
  arXiv:1206.1083 [hep-th].

  
\bibitem{Bekenstein:1992pj} 
  J.~D.~Bekenstein,
  ``The Relation between physical and gravitational geometry,''
  Phys.\ Rev.\ D {\bf 48}, 3641 (1993)
  [gr-qc/9211017].


  
  \bibitem{Schlamminger:2007ht} 
  S.~Schlamminger, K.~-Y.~Choi, T.~A.~Wagner, J.~H.~Gundlach and E.~G.~Adelberger,
  ``Test of the equivalence principle using a rotating torsion balance,''
  Phys.\ Rev.\ Lett.\  {\bf 100}, 041101 (2008)
  [arXiv:0712.0607 [gr-qc]].
  

  
  \bibitem{constr1} 
  B.~-A.~Gradwohl and J.~A.~Frieman,
  ``Dark matter, long range forces, and large scale structure,''
  Astrophys.\ J.\  {\bf 398}, 407 (1992).
  
  \bibitem{constr2} 
  R.~Bean, E.~E.~Flanagan, I.~Laszlo and M.~Trodden,
  ``Constraining Interactions in Cosmology's Dark Sector,''
  Phys.\ Rev.\ D {\bf 78}, 123514 (2008)
  [arXiv:0808.1105 [astro-ph]].








\end{thebibliography}
\end{document}